\documentclass[aps,prb,twocolumn,letterpaper]{revtex4}
\usepackage{graphicx}
\usepackage{dcolumn}
\usepackage{amsmath}
\usepackage{amssymb}
\usepackage{subfig}
\usepackage{graphicx}
\usepackage[countmax]{subfloat}
\usepackage{verbatim}
\usepackage{feynmf, tikz}

\usepackage{slashed}
\usepackage{color}
\def\Qv{{\bf Q}}

\def\kv{{\bf k}}

\def\pv{{\bf p}}

\def\beq{\begin{equation}}
\def\eeq{\end{equation}}
\def\beqa{\begin{eqnarray}}
\def\eeqa{\end{eqnarray}}

\begin{document}
\title{Zeeman field induced non-trivial topology in a spin-orbit coupled superconductor}
\author{Aaron Farrell and T. Pereg-Barnea}
\affiliation{Department of Physics and Center for the Physics of Materials, McGill University,
Montreal, QC, Canada}
\begin{abstract}
The hope to realize Majorana fermions at the vortex core of a two dimensional topological superconductor has led to a variety of proposals for devices which exhibit topological superconductivity.  Many of these include superconductivity through the proximity effect and therefore require a layer of a conventional superconductor deposited on top of another system, which lends its topological properties\cite{Fu,Sau,Alicea}.  The necessity of the superconducting layer poses some technical complications and, in particular, makes it harder to probe the Majorana state.  In this work we propose to replace the proximity effect pairing by an innate tendency for pairing, mediated by interactions.  We use a model system with spin orbit coupling and on-site repulsion and apply renormalization group to the interaction vertex.  Without a Zeeman field this model exhibits pairing instabilities in different channels depending on the tuning of parameters.  Once a Zeeman field is introduced the model favors topological superconductivity where the order parameter winds an odd number of times around the Fermi surface.  This suggests that certain superconductors, with strong spin-orbit coupling, may go through a topological phase transition as a function of applied magnetic field.
\end{abstract}
\maketitle

\section{Introduction}
Majorana fermions, interesting in their own right, are desirable components of topological, fault tolerant, quantum computations.  In order to perform such computations it is necessary to move Majorana fermions and in particular exchange their position in a braiding fashion.  While it is probably easiest to achieve Majorana fermions in one dimensional systems\cite{Mourik,Das}, controlling their motion seems more natural in two dimensions.  This reason and others inspire the search for two dimensional topological superconductors which are known to support Majorana fermions at their vortex cores\cite{Read, Gurarie}.

Some of the prominent ideas for two dimensional topological superconductivity include a multi-layer heterostructure\cite{Fu,Alicea,Sau,lutchyn,oreg,Cook,klinovaja}.  In the proposed structures one or more layers provide the topological properties, i.e, a winding of the electron spin around the Brillouin zone while another layer provides the tendency for pairing through the proximity effect.  For example, in a heterostructure of spin-orbit coupled semiconductors in proximity to a simple $s$-wave superconductor topological superconductivity arises as the pairing gap inherits the winding of the spins and forms a $p_x+ip_y$ state.  Besides SOC and pairing, a key ingredient in the above proposal is a Zeeman field.  The importance of the Zeeman field is in ensuring that only one Fermi surface with spin-momentum locking participates in the pairing.  Without the Zeeman field there are two spin-orbit coupled bands with opposite spin chirality in each energy.  This leads to an overall cancelation of the topology which is manifested in a trivial $Z_2$ invariant.

Inspired by the above proposals we set out to answer the following question.  Can the combined effect of spin-orbit coupling and electron-electron interaction lead to topological superconductivity?  Our study suggests that the answer is affirmative with the help of a Zeeman field.  Similarly to the semiconductor proposals, in our system the Zeeman field ensures that only one spin orbit coupled band is paired, producing a topological superconductor.  We therefore speculate that there may exist a spin-orbit coupled superconductor whose topology is trivial due to the multiplicity of Fermi surfaces\cite{wang1,wang2}.  This superconductor can be rendered topological by the application of a magnetic field.  Of course one may worry that the magnetic field has an orbital effect which ultimately leads to the suppression of superconductivity.  We therefore look at interaction driven superconductors which has the potential for a high critical field such that a topological superconductor phase may appear {\it before} superconductivity is completely turned off.

The model we use is an extension of the Hubbard model on the square lattice with Rashba spin orbit coupling (SOC).  Without SOC this model leads to a $d$-wave superconductor when treated in the strong coupling limit away from half filling\cite{Scalapino,Scalapino1,Corboz}.  In the presence of SOC coupling there are various phases depending on parameters.  While we have recently analyzed a similar model in  weak\cite{Farrell} and strong coupling\cite{FarrellSC}, in this paper we focus on its continuum analogue and find the possible pairing channels in an RG analysis. We find that when the Fermi level cuts only one of the spin-orbit coupled bands the interaction induces topological superconductivity.  This superconductor is of either $f$-wave or $p$-wave symmetry, depending on the direction of the Zeeman field relative to the spin winding.  This type of pairing comes about as a combination of two effects. Without SOC the preferred channel of pairing is a spin-singlet $d$-wave.  The SOC couples the two spin directions and produces two bands in which the spin winds by $2\pi$ as one encircles the Brillouin zone mid point.  This winding is superimposed on the $\pm 4\pi$ phase winding of the $d$-wave order parameter and leads to either a $2\pi$ ($p$-wave) or a $6\pi$ ($f$-wave) winding of the order parameter as seen by the band electrons.  Our renormalization group analysis shows that the above topological channels are dominant when there is a single Fermi surface.

This paper is laid out as follows. In Section \ref{sec:discuss} we discuss and review past heterostructure devices. In particular, we highlight the necessity for a description of the problem in a band basis. In this case the wave function of Cooper pairs is a mixture of singlet and triplet pairing\cite{GorkovRashba}. This section aims to frame the results we present in relation to the current literature on this problem.  It also gives an overview of our main results. Section \ref{sec:Model} defines the model we study and reviews the basics of the RG method we will employ. We then begin discussing our results in Section \ref{sec:RGresults} by first looking at the general problem and then specializing to systems with a single Fermi surface. After finding the potential for topological superconductivity in Section \ref{sec:RGresults} we follow up in Section \ref{sec:topo} by developing a simple mean-field model from our RG results and showing that for the correct combinations of parameters we get a topological superconductor. We close the main text in Section \ref{sec:conclusion} with some concluding remarks. A detailed appendix gives an overview of more technical details of our RG analysis.

\section{Preliminary Discussion}\label{sec:discuss}

\subsection{Superconductivity with lifted spin degeneracy}

As discussed in the introduction the goal of this paper is to address interaction driven superconductivity in a system of spin-orbit coupled electrons. We would first like to acquaint the reader with superconductivity in systems with a ``lifted spin degeneracy". Without spin-orbit coupling (and Zeeman field) the model we will use in this work simplifies to a tight binding model with spin-degenerate bands. Such a dispersion is typical in the studies of superconductivity with which most readers are familiar. In this case it is natural to discuss the formulation of Cooper pairs in singlet or triplet spin configurations.

Once spin-orbit coupling is introduced the two-fold spin-degeneracy of the bands is lifted and singlet and triplet pairings become mixed in the wave function of the Cooper pairs\cite{GorkovRashba}. Put another way, once spin-orbit coupling is considered the $z$ projection of the electron spins is no longer a good quantum number to describe the system with. Instead we are left with what we will be referred to as a band index. We can then see Cooper pairs form between two electrons in the same band (intraband pairing) or between electrons in different bands (interband pairing). These two types of pairs can, of course, be thought of as superpositions of the more traditional singlet and triplet pairs.

A clear illustration of this was given by Alicea in Ref.~[\onlinecite{Alicea}]. If one considers a system of spin-orbit coupled electrons, in this case a quantum well system, placed in proximity to an $s$-wave (singlet) superconductor an interpretation of such a system in terms of interband and intraband pairs is as follows. Proximity effect {\em forces} the electrons to pair in a spin-singlet state, symbolically we can think of this as adding a term $\Delta_0 \int d\kv \left(\psi^\dagger_{\uparrow}(\kv)\psi^\dagger_{\downarrow}(-\kv)+\text{h.c.}\right)$ to the Hamiltonian. However, in the quantum well the states $(\kv,\sigma)$ (where $\sigma=  \uparrow$ or $\downarrow$) are no longer good states and one must instead describe the system in a ``band" basis. Quantitatively this amounts to transforming the operators $\psi^\dagger_{\sigma}(\kv) $ into operators creating/destroying electrons in each band which we denote as $\psi_{\pm}(\kv)$. Carrying out this transformation the superconducting contribution to the Hamiltonian becomes (schematically)
\beqa
&&H_{SC}= \int d\kv (\Delta_{++}(\kv)\psi^\dagger_{+}(\kv)\psi^\dagger_{+}(-\kv)\\ \nonumber &+&\Delta_{--}(\kv)\psi^\dagger_{-}(\kv)\psi^\dagger_{-}(-\kv)+\Delta_{+-}(\kv)\psi^\dagger_{+}(\kv)\psi^\dagger_{-}(-\kv)+\text{h.c.})
\eeqa
The functions $\Delta_{++}(\kv), \Delta_{--}(\kv)$, and $\Delta_{+-}(\kv)$ play the role of the superconducting order parameter for, respectively, intraband pairing between two electrons in the upper band, two electrons in the lower band, and inter band pairing between one electron in the upper band and one in the lower band.

One additional, important observation when dealing with superconductivity with lifted spin degeneracy is that the symmetry of the pairing in the spin basis is not generally the same as the symmetry in the band basis. For the example above, Alicea begins with a system that has simple $s$-wave pairing in the spin basis; however, this pairing drives band Cooper pairs to form with the order parameters $\Delta_{++}(\kv), \Delta_{--}(\kv)$, and $\Delta_{+-}(\kv)$ which are non-trivial functions of $\kv$.  In particular, $\Delta_{++}(\kv)$ and $\Delta_{--}(\kv)$ are odd under $\kv\to-\kv$ while $\Delta_{+-}(\kv)$ is even.

\subsection{Spinless $p$-wave Pairing in the band basis}

Superconductivity with $p\pm ip$ pairing is highly desirable as it is the canonical example of a topological superconductor\cite{Read}. The example above can be argued to be a spinless $p-ip$ superconductor. In order to review this argument let us first quickly review the band structure of these systems. One begins with two, spin degenerate, quadratic bands. When spin-orbit coupling is added these bands are split and the degeneracy is lifted. If the bands are parabolic they cross at $\kv=0$. When a Zeeman field is applied the crossing is avoided and a gap is opened around $\kv=0$.  Therefore, for any energy within this $\kv=0$ gap there is a single circular contour of constant energy on which the spin is locked to the momentum direction. Please note that when we discuss a Zeeman field opening a gap in the rest of this paper we are referring to this scenario. An example of this type of band structure can be seen in Fig.~\ref{fig:schematic}.   If the chemical potential of the system is tuned so that the Fermi surface lies in this gap, and superconductivity is not strong enough to induce transitions between bands (through $\Delta_{+-}(\kv)$) then the upper band of the problem plays no role and can be projected out. This leaves only a Hamiltonian
\beqa
&&H_{eff} =\int d\kv (\epsilon_{-}(\kv)\psi^\dagger_{-}(\kv)\psi_{-}\\ \nonumber &+&\Delta_{--}(\kv)\psi^\dagger_{-}(\kv)\psi^\dagger_{-}(-\kv)+\Delta_{--}^*(\kv)\psi_{-}(-\kv)\psi_{-}(\kv))
\eeqa
 where $\epsilon_{-}(\kv)$ is some dispersion. This is exactly a spinless $p-ip$ superconductor.

While the above idea was proposed for a system with $s$-wave pairing the same principle can be applied for other singlet superconductors. Of interest for the current work is that $d+id$-wave singlet pairing leads to $f+if$ and $p+ip$ intraband pairing which are both topologically non-trivial\cite{Farrell, Sato1}.

The approach taken in this paper, although highly motivated by the above discussion, works in the opposite direction. Instead of inducing pairing {\em via} proximity effect we look at driving pairing by interactions. Rather than forcing pairs to develop in, say, the $s$-wave singlet channel as the heterostructure devices above do, we utilize renormalization group methods to look at instability for pairing between band electrons. We focus on the topologically relevant band structure discussed above, that is not just a spin-orbit split band structure but one with a gap opened {\em via} a some sort of mass (Zeeman) term.

It has been shown in Refs.~[\onlinecite{wang1, wang2}] that spin-orbit coupling in an otherwise quadratic band structure leads to enhancement of superconductivity. These works find an instability towards pairing with the symmetry of the order parameter (or at least the dominant term) dependent on how the relative strength of the spin-orbit coupling and fermi energy are tuned. Here we follow a similar program but with the introduction of a mass term into the model. This term opens a gap between the spin-orbit split bands in the {\em non-interacting} band structure and we focus on what happens when the chemical potential is tuned to lie in the gap. For this choice of parameters there is only a single Fermi surface and we find that the pairing that develops has either $p+ip$ or $f+if$ symmetry depending on the sign of the Zeeman mass term. In either case we expect the superconductivity that develops to be topological in nature, i.e., to support Majorana fermions in its vortex cores.

\section{Model and Method}\label{sec:Model}
\subsection{Model}

\subsubsection{Definition}
Here we would like to introduce our model and the language of a band basis that the rest of this work will be framed in. Our initial focus is on the Hamiltonian studied in [\onlinecite{Farrell}] and [\onlinecite{FarrellSC}]:
\beq
H = H_1 + H_{int}
\eeq
where $H_1=H_{KE}+H_{SO}$ is a quadratic Hamiltonian and $H_{int}$ contains interactions effects. For $H_1$ we take the following model
\beq
H_1 = \sum_{\kv, \alpha, \beta} c_{\kv, \alpha}^\dagger \left(\xi_\kv \delta_{\alpha,\beta } + {\bf d}_\kv \cdot \vec{{\bf \sigma}}_{\alpha,\beta}\right) c_{\kv, \beta}
\eeq
where $\xi_\kv = \epsilon_\kv -\tilde{\mu}$ with $\epsilon_\kv = -2t(\cos(k_x)+\cos(k_y))$ and ${\bf d}_\kv = (A\sin{k_x}, A\sin{k_y}, 2B(\cos{k_x}+\cos{k_y}-2)+M)$ where $t,A,B,M,\tilde{\mu}$ are material parameters giving the strength of the hopping amplitude, in plane spin-orbit coupling, out-of-plane spin-orbit coupling, Zeeman field and chemical potential respectively. Above $\vec{\sigma}$ is a vector of Pauli matrices and $\alpha$ and $\beta$ are spin labels.

The model above has been chosen for the sake of versatility. We have discussed numerous possible applications in past work\cite{FarrellSpin}, they include cold atomic systems with synthetic gauge fields, transition-metal oxides ({\em e.g.} pyrochlore iridates), quantum wells, and insulating oxide interfaces ({\it e.g.} the interface of LaAlO$_3$ and SrTiO$_3$). From this point of view the parameters $A$ and $B$ could come from traditional spin-orbit coupling such as Rashba or Dresselhaus,  they could find their origins in systems like quantum wells\cite{BHZ}, or they could be created in a cold atomic system. At the same time $M$ could come from applied field, proximity to a FM insulator or a band gap.

To account for interactions we take a simple on-site coulomb repulsion given by
 \beqa
H_{int} &=& \sum_{\kv_1, \kv_2, \kv_3, \kv_4}\sum_{\alpha_1,\alpha_2,\alpha_3,\alpha_4}\delta_{\kv_1+\kv_2,\kv_3+\kv_4}\\ \nonumber &\times& {U}_{\alpha_1,\alpha_2,\alpha_3,\alpha_4}(\kv_1, \kv_2, \kv_3, \kv_4) c_{\kv_1, \alpha_1}^\dagger c_{\kv_2, \alpha_2}^\dagger  c_{\kv_3, \alpha_3} c_{\kv_4, \alpha_4}
\eeqa
 where
\beqa
&&{U}_{\alpha_1,\alpha_2,\alpha_3,\alpha_4}(\kv_1, \kv_2, \kv_3, \kv_4) = \\ \nonumber && \frac{U}{4N} \left( \sigma^x_{\alpha_1,\alpha_2}\delta_{\alpha_1, \alpha_4}\delta_{\alpha_2, \alpha_3}-  \sigma^x_{\alpha_2,\alpha_1}\delta_{\alpha_2, \alpha_4}\delta_{\alpha_1, \alpha_3}\right).
\eeqa

Note that it is enough to take a {\em repulsive} interaction since it leads to pairing in a strong coupling treatment (in contrast to Ref.~[{\onlinecite{Farrell}] where near-neighbor attraction was introduced to mimic this effect in the weak coupling treatment).

\subsubsection{Transformation to the Band Basis}

We now diagonalize $H_1$ in order to recast our problem in terms of band electrons. This is done by making the unitary transformation
 \beq \label{unitary}
   \left( \begin{matrix} 
       c_{\kv,\uparrow} \\
       c_{\kv, \downarrow} \\
    \end{matrix}\right) = \left(   \begin{matrix} 
          f_{+1}(\kv) & f_{-1}(\kv) \\
          e^{i\theta_{\kv}} f_{-1} (\kv)& - e^{i\theta_{\kv}} f_{+1}(\kv)  \\
       \end{matrix}\right)  \left( \begin{matrix} 
       b_{\kv,+} \\
       b_{\kv, -} \\
    \end{matrix}\right)
 \eeq
 where the $+1, -1$ label a band and we have defined the following
 \beqa
 e^{i\theta_{\kv}} &=& \frac{d_1(\kv)+id_{2}(\kv)}{\sqrt{d_1(\kv)^2+d_2(\kv)^2}} \\ \nonumber
 f_{\lambda} (\kv)&=& \sqrt{\frac{d+\lambda d_3}{2d}}
 \eeqa
where $\lambda=\pm1$ and $d=|{\bf{d_k}}|$. Written in the new basis,
\beq
H_1 = \sum_{\kv, \lambda} E_{\kv,\lambda} b_{\kv,\lambda}^\dagger b_{\kv,\lambda}
\eeq
where $b_{\kv\lambda}$ are annihilation operators in the band $\lambda$ and $E_{\kv,\lambda} = \xi_\kv +\lambda d$ labels the energy of the bands relative to the Fermi energy $\mu$.

Before we move on to the RG calculations we express $H_{int}$ in terms of band electrons. This requires some tedious manipulation which we defer to the Appendix. After some work we obtain:
\beqa
H_{int} &=&\sum_{\kv_1, \kv_2, \kv_3, \kv_4}\sum_{\lambda_1,\lambda_2,\lambda_3,\lambda_4}\delta_{\kv_1+\kv_2,\kv_3+\kv_4}\\ \nonumber &\times& {W}_{\lambda_4,\lambda_3,\lambda_2,\lambda_1}(\kv_4, \kv_3, \kv_2, \kv_1)   b_{\kv_4, \lambda_4}^\dagger b_{\kv_3, \lambda_3}^\dagger  b_{\kv_2, \lambda_2} b_{\kv_1, \lambda_1}
\eeqa
where
\beqa
{W}_{\lambda_4,\lambda_3,\lambda_2,\lambda_1}(\kv_4, \kv_3, \kv_2, \kv_1)  &=& -\frac{U}{4N}w_{\lambda_4, \lambda_3}(\kv_4, \kv_3) \nonumber\\ &\times& w^*_{\lambda_2, \lambda_1}(\kv_2, \kv_1)
\eeqa
where $w_{\lambda_i, \lambda_j}(\kv_i, \kv_j)=\lambda_i e^{-i\theta_{\kv_i}}F_{j,i}-\lambda_j e^{-i\theta_{\kv_j}}F_{i,j}$ with $F_{i,j} =  f_{\lambda_i} (\kv_i) f_{-\lambda_j} (\kv_j) $. We see that in the band basis electrons in bands $\lambda_2$ and $\lambda_1$ can scatter to bands $\lambda_3$ and $\lambda_4$, that is to say there is no ``band conserving condition". The $\delta$-function ensures that the momentum is conserved.

\subsubsection{Partition Function}

We can now recast our model in the language of coherent state path integral. First we define the quadratic part of the action as
\beq
S_0 = \int_0^\beta d\tau  \sum_{\kv, \lambda} b_{\kv,\lambda}^*(\tau) \left(\frac{\partial}{\partial \tau} + E_{\kv,\lambda} \right) b_{\kv,\lambda}(\tau)
\eeq
with Grassman variables $b_{\kv,\lambda}^*(\tau) $ and $b_{\kv,\lambda}(\tau) $. The interaction part is:
\beq
S_{int} = \sum_{1,2,3,4} { V} (4,3,2,1) b^*(4)b^*(3)b(2)b(1)
\eeq
where we simplified the notation by defining $i=(\lambda_i, \kv_i, \tau_i)$ and
\beqa
{ V}(4,3,2,1) &=& -\frac{U}{4N} \int_0^\beta  d\tau \left(\prod_{j=1}^4\delta(\tau-\tau_j)\right) \delta_{\kv_1+\kv_2,\kv_3+\kv_4}   \nonumber\\ &\times& w_{\lambda_4, \lambda_3}(\kv_4, \kv_3) w^*_{\lambda_2, \lambda_1}(\kv_2, \kv_1).
\eeqa
The partition function is given by:
\beq
Z = \int \mathcal{D}(b^*_{\lambda}(\tau), b_{\lambda}(\tau))e^{-S_0-S_{int}}
\eeq

\subsection{Renormalization Group Approach}

We now take the standard steps in finding the renormalization group flow of our model\cite{shankar}. We begin by separating the Grassman variables into fast and slow modes:
 \beqa
 b_{\kv, \lambda}(\tau)& =&  \theta(\Lambda/s -|E_{\kv, \lambda}|)b_{\kv, \lambda}^{<}(\tau) \\ \nonumber &+&\theta(\Lambda -|E_{\kv, \lambda}|)\theta( |E_{\kv, \lambda}|- \Lambda/s) b_{\kv, \lambda}^{>}(\tau)
 \eeqa
 where $\Lambda$ is our energy cut-off and $s$ is a flow-parameter. While fast and slow modes are decoupled in $S_0$, they are coupled in $S_{int}$
 \beq
 S = S_0 +S_{int} = S_0(<)+S_0(>)+S_{int}(<,>)
 \eeq
Integrating over all of the fast modes gives
 \beqa
Z &=&\int \mathcal{D}(b_<, b_<^*) e^{-S_0(<)-S'_{int}(<)}
\eeqa
 where $-S'_{int}(<) = \ln\left[\langle e^{-S_{int}(<,>)} \rangle_{0,>}\right]$ where the average is over fast modes with respect to $e^{-S_0(>)}$. We can obtain an approximation of $S'_{int}$ by performing a cumulant expansion:
 \beqa
-S'_{int}(>) &=& -\langle  S_{int}(<,>) \rangle_{0,>} \\ \nonumber &+& \frac{1}{2}\left(\langle S_{int}^2(<,>) \rangle_{0,>}-\langle S_{int}(<,>) \rangle_{0,>}^2\right)  \\ \nonumber &-& \frac{1}{3!}\langle\langle S_{int}^3(<,>) \rangle\rangle_{0,>} +\frac{1}{4!}\langle\langle S_{int}^4(<,>) \rangle\rangle_{0,>}
 \eeqa
 where the double angled brackets denote, respectively, the third and fourth order cumulants of $S_{int}$ (with respect to $e^{-S_0(>)}$). Using the above perturbative expression we calculate a new effective interaction for the slow modes of the theory. We would like to point out that all of the results we derive are perturbative {\em only} in the interaction strength $U$, while $t$, $A$, $B$, $M$ and $\tilde{\mu}$ are not assumed small in any way.

 We use Feynman diagrams in order to evaluate the above expression\cite{shankar}. The relevant Feynman rules are as follows:
 \begin{itemize}
 \item Each vertex diagram contains 4 external lines, two incoming and two outgoing. All other lines will be referred to as internal.
  \item Label every line with a momentum $\kv$, a band index $\lambda$ and a Matsubara frequency $i\omega_m$.
  \item For every internal line write a bare propagator $G_{\kv,\lambda}(i\omega_m)=\frac{1}{i\omega_m - E_{\kv,\lambda}} $.
  \item Every vertex in the diagram has a factor of $ {V}_{\lambda_4,\lambda_3,\lambda_2,\lambda_1}(\kv_4, \kv_3, \kv_2, \kv_1)=
    - \frac{U \left(  \lambda_4e^{-i\theta_{\kv_4} }  F_{3,4}- \lambda_3e^{-i\theta_{\kv_3} }F_{4,3}  \right)\left(  \lambda_2e^{i\theta_{\kv_2} } F_{1,2} -\lambda_1e^{i\theta_{\kv_1} }  F_{2,1}  \right) }{4N}$ where the numbers 1 through 4 {\em must} be assigned in the following way: For propagators coming into the vertex, the one from the left is 1 and the one from the right is 2, for propagators leaving the vertex the one from the left is 3 and from the right is 4.
    \item Conserve total momenta and frequency at each vertex.
    \item Sum over all internal frequencies $\frac{1}{\beta} \sum_{i\omega_m}$.
    \item Sum over all internal momenta, with the restriction that these are fast modes $ \sum_{\kv, >}$.
    \item Sum over all internal band indices.
    \item Determine the overall multiplicative factor by multiplying by how many independent ways there are of drawing a specific diagram and for n$^{th}$ order diagrams divide by $n!$.
    \item Multiply by the appropriate sign in the cumulant expansion and determine the sign coming from the contractions required to draw a given diagram
  \end{itemize}

Once the above steps have been completed we have a {\em general} expansion. We focus on the ``BCS" (pairing) channel by setting  the incoming external momenta to $\pm\kv'$  and the outgoing external momenta to $\pm\kv$ and band indices to $\lambda$. Further, here we are interested in only intraband pairing and so we set both incoming external band indices to $\lambda'$ and the outgoing ones to $\lambda$. The calculation is also further simplified by only considering diagrams that contain logarithmic divergences as $s\to\infty$\cite{wang1,wang2}, as these dominate the vertex function.

We focus only on intraband pairing both because it is the interesting pairing from a topological stand-point and because we expect interband pairing to be suppressed relative to its interband counterpart. For example, for the BCS diagram (see the appendix of this paper) we find a relative suppression of $\left(\frac{M}{Ak_f}\right)^2$ in the interband pairing.

 \section{Summary of RG Analysis}\label{sec:RGresults}

 \subsection{General Results}
 Here we discuss the results of our analysis. We will only include details where it is absolutely essential; a full treatment and discussion of our analysis is included in the appendix.

 We have calculated diagrams up to fourth order in the interaction parameter $U$ and with logarithmic divergences for our lattice model. Although we have successfully obtained an expression of the renormalized interaction up to this order it is rather intractable to work with this form. In order to make analytic progress we take the continuum limit by replacing $\sin{k_i}\to k_i$, $\cos{k_i} \to 1-k_i^2/2$.

 Taking the continuum limit not only makes our mathematical expressions analytically tractable but also makes the dispersion a function of $k=|\kv|$ only, such that the contours of constant $E_{\kv,\lambda}$ are circular in $\kv$-space. This enables two simplifications\cite{shankar, wang2}: (1) We can set all external momenta to lie on the Fermi surface (or Fermi surfaces) as any other external momenta correspond to processes that are irrelevant under RG flow and (2) due to the circular symmetry the coupling function in the BCS channel can be a function of $\phi= \theta_{\kv}-\theta_{\kv'}$ only.

 When considering the BCS channel we can think of the RG procedure as renormalizing the interaction as $H_{int}^{eff} = \sum_{\kv,\kv', \lambda, \lambda'} V'_{\lambda,\lambda'}(\kv,\kv', s)  c^\dagger _{\kv,\lambda}c^\dagger_{-\kv,\lambda} c_{-\kv',\lambda'}  c_{\kv',\lambda'}$.   In the continuum limit the interaction takes the form
 \beqa
 V'_{\lambda,\lambda'}(\kv,\kv', s)  \equiv V'_{\lambda,\lambda'}(\phi, s) =\frac{ e^{i\phi}}{N}v'_{\lambda,\lambda'}(\phi, s)
\eeqa
Where we have used primed variables to distinguish renormalized parameters from the non-primed bare variables.
The indices $\lambda$ and $\lambda'$ label the Fermi surface of the outgoing and incoming band electrons.

 Using  this symmetry of the coupling allows us to decouple the interaction into angular momentum channels:
\beqa\label{angmom}
v'_{\lambda,\lambda'}(\phi, s)  &=& \sum_{m_z} e^{im_z\phi} v'_{\lambda,\lambda'}(m_z,s) \\ \nonumber
 v'_{\lambda,\lambda'}( m_z, s) &=& \frac{1}{2\pi} \int_0^{2\pi}e^{-im_z\phi} v'_{\lambda,\lambda'}(\phi, s)
\eeqa

We now address the question of how does the function $v'_{\lambda,\lambda'}( m_z, s) $ flow as the RG parameter $s$ is tuned and calculate its beta-function. In our full expression for  $v'_{\lambda,\lambda'}( m_z, s)$ (see the appendix of this paper) the matrix entries for different $\lambda$'s are coupled. Fortunately we can obtain a simple beta function by following a method proposed by Raghu {\it et al}\cite{raghu1} and further employed by Vafek and Wang\cite{wang1, wang2}. To this end we define the $g'$ matrix
\beqa
g'_{\nu,\mu}(s,m_z) = \sqrt{N_{\nu} N_{\mu}}v'_{\nu, \mu}(s, m_z)
\eeqa
In this definition $N_{\mu}$ is the density of states at the Fermi energy for the band $\mu$. We then obtain the following beta function for the eigenvalues of each $g'$-matrix
\beqa
\frac{d \lambda^{m_z}_i(s)}{d \ln(s)}  = -2   (\lambda^{m_z}_i(s))^2
\eeqa
 where $ \lambda^{m_z}_i(s)$ is the $i^{th}$ eigenvalue of $g'_{\nu,\mu}(s,m_z)$. With the above beta-function,
\beq
 \lambda^{m_z}_i(s) = \frac{ \lambda^{m_z}_i(1)}{1+2 \lambda^{m_z}_i(1)\ln(s)}.
\eeq
From this solution we see that if $\lambda^{m_z}_i(1) <0$ the renormalized coupling diverges at $s=e^{-\frac{1}{2 \lambda^{m_z}_i(1)}}$.  Thus for any $i$ or $m_z$, if $\lambda^{m_z}_i(1) <0$ superconductivity will develop in this channel\cite{wang2} with a superposition of intraband pairing given by the eigenvector corresponding to $\lambda^{m_z}_i(1)$. The temperature scale at which it will develop is given by the value of the lower cut-off $\Lambda/s$ at which the above solution diverges, namely \cite{raghu1, wang1, wang2} $T_c \sim \Lambda e^{\frac{1}{2 \lambda^{m_z}_i(1)}}$. Although this argument does not allow us to quantitatively determine $T_c$ due to the lack of a proportionality constant, it does allow us to compare the transition temperatures of different channels ($i$ and $m_z$) for a given set of parameters. The more negative the eigenvalue $\lambda^{m_z}_i(1)$ the higher $T_c$ will be. Therefore we can think of $\lambda^{m_z}_i(1) $ as a measure of the instability of a particular channel. For a given set of parameters, the value of $m_z$ and $i$ with the most negative $\lambda^{m_z}_i(1)$ is the dominant superconductivity channel.

The above discussion is general for intraband pairing for any choice of parameters in the model. Let us now specialize to the case of interest, a single Fermi surface.

\subsection{Pairing on a Single Fermi Surface}

First let us describe the two band dispersion $\epsilon_{\kv, \pm1}$.  The upper band $\lambda=+1$ is a monotonically increasing function of $k$ and parabolic like. For $\lambda=-1$ there are two possible functional forms depending on parameters. The first is a parabolic-like, monotonically increasing function of $k$. The second has a Mexican hat shape with a value of $-|M|$ at $k=0$ and a minimum at some finite momentum $k =k_{min}$. Above this value of $k$ the band energy is monotonically increasing. A schematic plot of $\epsilon_{\kv,\pm1}$ is shown in Fig.~\ref{fig:schematic}.

 \begin{figure}[tb]
  \setlength{\unitlength}{1mm}
\vspace{5mm}
   \includegraphics[scale=.4]{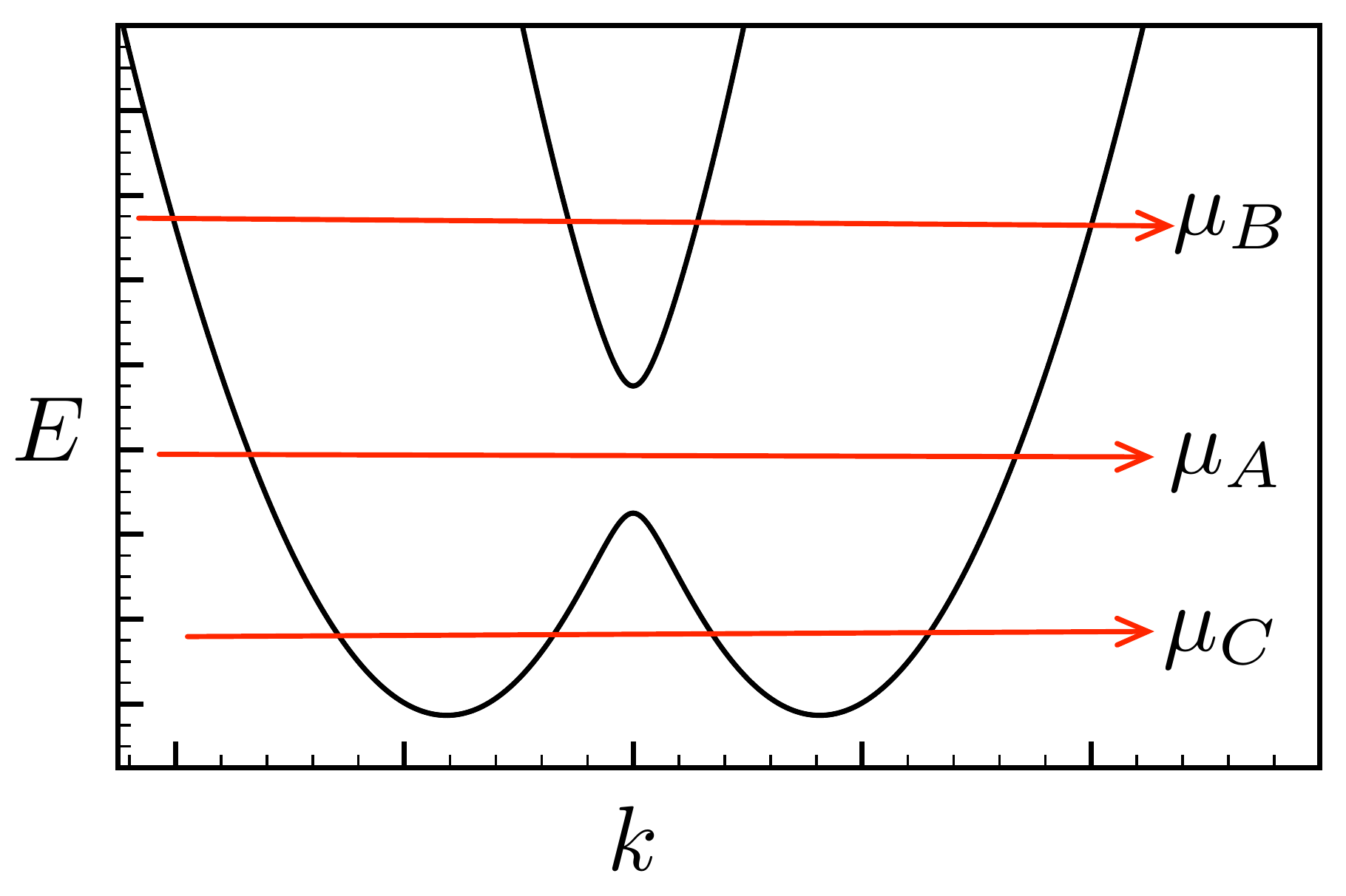}
\caption{{\small
Schematic picture of our linearized bands. The different horizontal lines correspond to values of the chemical potential creating a single Fermi surface ($\mu_A$) and to two Fermi surfaces ($\mu_{B}$ and $\mu_C$).
     }
     }\label{fig:schematic}
\end{figure}

Regardless of the functional form of $\epsilon_{\kv,-1}$ the two bands are separated at $k=0$ by an energy $2|M|$. The interesting regime is when $|\mu|<|M|$. In this regime there is a single fermi surface, marked by the line $\mu_A$ in Fig.~\ref{fig:schematic}.  For $|\mu|>|M|$ we have either two  or zero Fermi surfaces; these regimes are marked by the lines $\mu_B$ and $\mu_C$ in Fig.~\ref{fig:schematic}.

Here we focus on the region $|\mu|<|M|$ where there is a single, circular Fermi surface with radius $k=k_F$. For this band structure slow modes must belong to the $\lambda=-1$ band. We thus focus on intra-band pairing of electrons in the $\lambda=-1$ band. This is the interacting analogue of the intraband pairing discussed in Refs.~[\onlinecite{Sau,Alicea}]. Note that our RG description naturally dispenses with interband pairing for this choice of parameters, {\it i.e} the Fermi surface contains only $\lambda=-1$ electrons and so we do not expect any interesting flow between electrons in opposite bands. In the language developed in Section \ref{sec:discuss} this would seem to suggest we would not expect to see $\Delta_{+-}(\kv)$ type pairings. Additionally, for similar reasons we do not expect to see $\Delta_{++}(\kv)$ type pairings either.

For this particular choice of parameters $N_{+1}=0$ as there are no states in the band $+1$ inside the Fermi surface.
The matrix $g$ reduces to a scalar which is obtained by setting $\lambda=\lambda'=-1$ and
 \beqa
\lambda^{m_z}(1) = \frac{U^2} {2^6}N_{-1} v_{-1,-1}(m_z).
 \eeqa

Where $v_{-1,-1}(m_z)$ is a numerical constant for a given set of material parameters and is defined in the appendix. From the above we see that the channel $m_z$ with the most negative value of $v_{-1,-1}(m_z)$ will be the dominant channel for pairing, as $\frac{U^2}{2^6}N_{-1}$ is a positive constant. The corresponding critical temperature can be roughly estimated as $T_c \sim e^{\frac{2^5}{U^2N_{-1} v_{-1,-1}(m_z)}}$.

As shown in the appendix  $v_{-1,-1}(m_z)$ is a complicated integral. Setting $B=0.0$ for both simplicity and to make a closer analogy with the work in [\onlinecite{Sau}], we have evaluated $v_{-1,-1}(m_z)$ numerically over the range of values, $A\le t$, $|M|\le.3t$ and $|\mu|< |M|$. For the parameters we have looked at, we find quite generally that for $M>0$ the dominant angular momentum channel is $m_z=2$ while for $M<0$ it is $m_z=-2$. As an example of the data we have plotted $v_{-1,-1}(m_z)$ as a function of $A$ in Fig.~\ref{fig:vm}. We note that in the absence of the mass term $M$ there is a degeneracy between $v_{-1,-1}(m_z)$ and $v_{-1,-1}(-m_z)$, quite simply $v_{-1,-1}(m_z)=v_{-1,-1}(-m_z)$ for all parameter values\cite{wang2}. This means that for $M=0$ the $m_z$ channel and the $-m_z$ channel are equally favorable. When we allow for a finite $M$ this degeneracy lifts.  For the choice of parameters in  Fig.~\ref{fig:vm} there is a large difference between the $m_z$ and $-m_z$ couplings and there is only a single dominant channel.

 \begin{figure}[tb]
  \setlength{\unitlength}{1mm}
\vspace{5mm}
   \includegraphics[scale=.45]{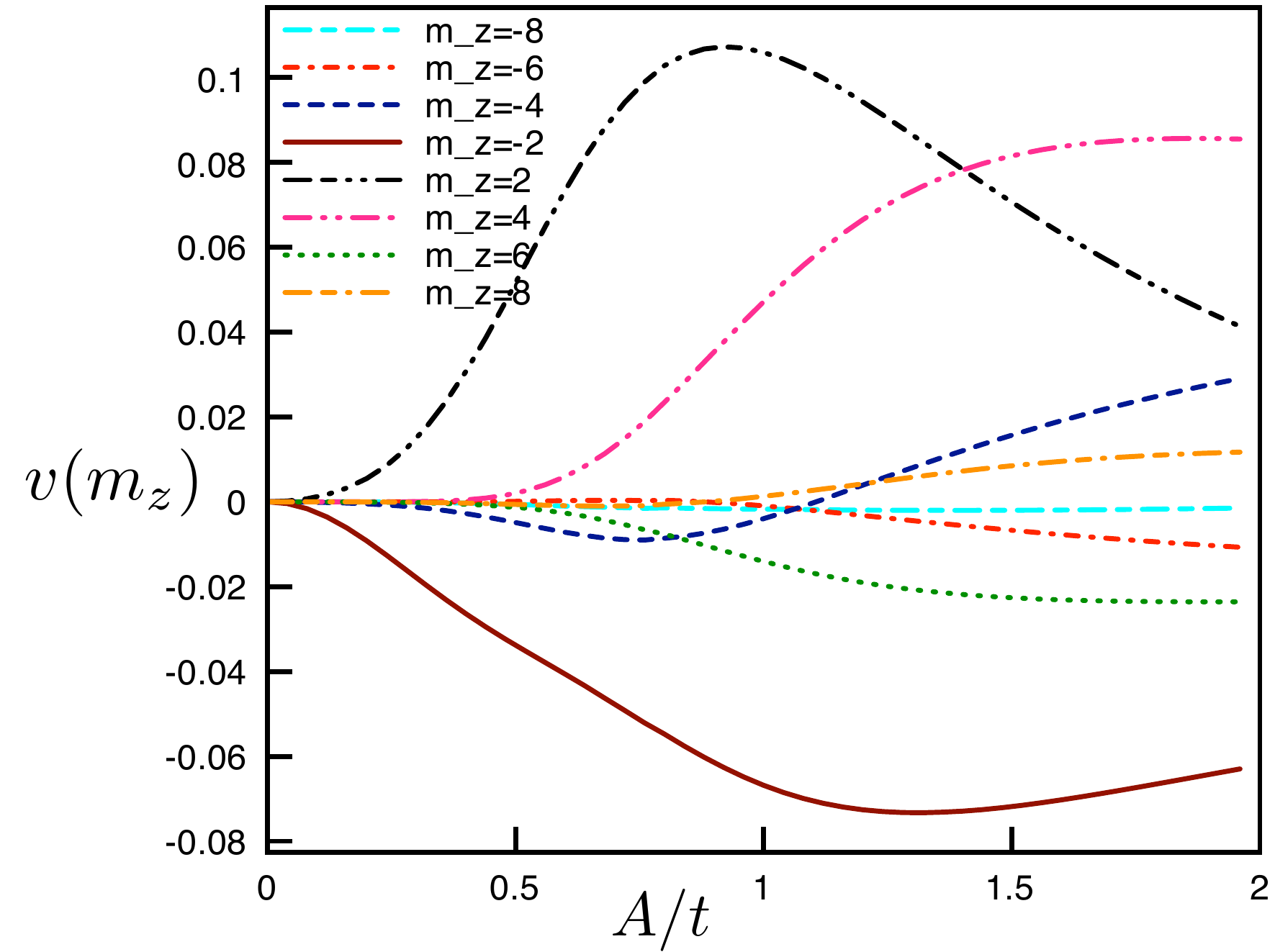}
\caption{{\small
Plot of the integral $v(m_z) = v_{-1,-1}(m_z)$ as a function of $A$. Above we have set $B=0$, $M=-0.3t$ and $\mu=0$. We note that for the range of parameters chosen the $m_z=-2$ channel is by far the most negative channel.
     }
     }\label{fig:vm}
\end{figure}

 Looking back at the original pairing function $V'_{\lambda,\lambda'}(\phi, s)= \frac{ e^{i\phi}}{N}v'_{\lambda,\lambda'}(\phi, s) $ and replacing $v'_{\lambda,\lambda'}(\phi, s) $ with its most dominant $m_z=\pm2$ component we see that for $|M|>0$ our RG analysis gives an attractive interaction with $f$-wave ($p$-wave) symmetry. Owing to the extra phase factor in this basis, this leads, in a mean field description (discussed in the next section) to an order parameter $\Delta_{--}(\kv)$ with $f$-wave or $p$-wave symmetry. In Section \ref{sec:discuss} we argued that a $p$-wave (and later an $f$-wave)  $\Delta_{--}(\kv)$  intraband order parameter with the chemical potential tuned in the gap should constitute a spinless topological superconductor. Here the same argument holds and we have found, within our analysis, that neither $\Delta_{++}(\kv)$ nor $\Delta_{+-}(\kv)$ type pairing develops.

Another encouraging comparison comes from connecting our RG calculation with the earlier mean-field work on the same model in  [\onlinecite{Farrell}].   The dominant $m_z=\pm2$, representing $f$-wave or $p$-wave band pairing leads to $d+id$-wave spin-singlet pairing and $p+ip$ and $f+if$-wave spin-triplet pairing when transformed back to the usual spin basis. Therefore, as we've seen in Section \ref{sec:discuss}, once we spin degeneracy is broken superconductivity may be a superpositions of spin-singlet and spin-triplet. The connection to Ref.~[\onlinecite{Farrell}] comes from noting that in this study $d+id$ singlet pairing was found using a variational mean field theory technique.  That study did not include $p$- or $f$- triplet pairing in the variational wave function and therefore those were not obtained.

\section{Mean Field Topological Classification} \label{sec:topo}

In order to gain more intuition and allow a simple evaluation of the topological invariant we resort to an effective mean field theory.  The RG analysis above points to the important component of the interaction and we therefore include only this component in the effective model.
To this end, we dispense with any interactions between $\lambda=+1$ band electrons and furthermore only consider the dominant $m_z=m_z^{opt}$ interaction channel between our $\lambda=-1$ band electrons. We therefore write the following dominant channel Hamiltonian,
\beqa\label{Heff}
H_{eff} &=& \sum_{\kv, \lambda} (\epsilon_\kv+\lambda d_{\kv}-\mu) b^\dagger_{\kv,\lambda}b_{\kv,\lambda}\\ \nonumber &+& \sum_{\kv,\kv'} V_{eff}(\phi) b^\dagger_{\kv,-1}b^\dagger_{-\kv,-1}b_{-\kv',-1}b_{\kv',-1}
\eeqa
where $V_{eff}(\phi)$ represents the part of the interaction which favors superconductivity in the dominant channel, denoted by $m_z \equiv m_z^{opt} = \pm 2$ and therefore $V_{eff}(\phi) \simeq \frac{e^{i(m_z^{opt}+1) \phi}}{N} v_{eff}(m_z^{opt})$. We perform a mean-field decoupling of the interaction term:
\beqa
&&H_{eff, MF} = \sum_{\kv, \lambda} \xi_{\kv,\lambda} b^\dagger_{\kv,\lambda}b_{\kv,\lambda} \\ \nonumber &+&\frac{1}{2} \sum_{\kv} \bigg{(}\Delta_0 e^{-i(m_z+1)\theta_{\kv}} b^\dagger_{\kv,-1}b^\dagger_{-\kv,-1} \\ \nonumber &+& \Delta_0 e^{i(m_z+1)\theta_{\kv}} b_{-\kv,-1} b_{\kv,-1} \bigg{)}
\eeqa
where we have dropped the `opt' superscript on $m_z$, defined $\xi_{\kv,\lambda}\equiv \epsilon_\kv+\lambda d_{\kv}-\mu$, and the parameter $\Delta_0$ must be determined self-consistently through the equation
\beq
1 = -\frac{v_{eff}(m_z^{opt})}{N}\sum_{\kv}\frac{\tanh\left(\frac{E_{\kv}}{2k_BT}\right)}{E_{\kv}}
\eeq
where $E_{\kv}=\sqrt{\xi_{\kv, -1}^2+\Delta_0^2}$. To obtain a description in terms of traditional spin states we transform $H_{eff, MF}$ back to a spin basis by inverting the transformation in Eq.~(\ref{unitary}) and arrive at
\beq
H_{eff,MF} = \frac{1}{2} \sum_{\kv} \psi_{\kv}^\dagger \mathcal{H}_{\kv} \psi_{\kv}
\eeq
where $ \psi_{\kv} = (c_{\kv,\uparrow}, c_{\kv, \downarrow}, c^\dagger_{-\kv, \downarrow}, -c^\dagger_{-\kv,\uparrow})^T$ and
\beq
\mathcal{H}_{\kv} =    \left(\begin{matrix} 
      h_{\kv} & \hat{\Delta}_{\kv} \\
      \hat{\Delta}^\dagger_{\kv} & -\sigma_y h_{-\kv}\sigma_y \\
   \end{matrix}\right)
\eeq
with $h_{\alpha,\beta}= (\epsilon_{\kv}-\mu)\delta_{\alpha,\beta} + {\bf d}(\kv)\cdot \vec{\sigma}_{\alpha,\beta}$ and
\beq
\hat{\Delta}_{\kv} =    \left(\begin{matrix} 
      \Delta_s& -\Delta_t^{\uparrow} \\
      \Delta_t^{\downarrow} &\Delta_s \\
   \end{matrix}\right)
\eeq
where
\beqa
\Delta_s &=&\frac{ \Delta_0}{2} \left(\frac{Ak}{d} \right)e^{-im_z\theta_{\kv}}\\
\Delta_t^{\uparrow} &=&  \frac{ \Delta_0}{2}\left(\frac{d-d_3}{d} \right) e^{-i(m_z+1)\theta_{\kv}}\\
\Delta_t^{\downarrow} &=& -\frac{ \Delta_0}{2}\left(\frac{d+d_3}{d}\right) e^{-i(m_z-1)\theta_{\kv}}
\eeqa
where $d$ and $d_3$ are functions of $k=|\kv|$. From this we can see that for $m_z=\pm 2$ we get $d+id$-wave singlet pairing and $f+if$ and $p+ip$-wave triplet pairing.

 We now calculate the Chern number of this effective mean field Hamiltonian. The BdG Hamiltonian $\mathcal{H}_\kv$ has particle-hole symmetry which greatly simplifies the evaluation as noted by Ghosh {\it et al}\cite{ghosh}. We define
  \beq
 Q(\kv) = -\text{sgn}\left(\text{Pf}(\mathcal{H}_\kv \Gamma)\right)
 \eeq
 where $\Gamma= \sigma_y \otimes \tau_y$ and `Pf' stands for the Pfaffian of the matrix argument.
 The Chern number is a function of $Q(\kv)$ evaluated at the time reversal invariant momenta (TRIM) and in the square lattice this amounts to
 \beq\label{chern1}
 C_1 = \frac{1}{i\pi} \ln\left[\frac{Q(0,0)Q(\pi,\pi)}{Q(\pi,0)Q(0,\pi)}\right]
 \eeq
 where in the logarithm we have taken a branch such that $\ln(-1)=i\pi$. We can easily calculate $Q$ and find
 \beq
 Q(\kv) = \text{sgn}\left(|\Delta_s(\kv)|^2+(\epsilon_\kv-\mu)^2-d_{\kv}^2\right)
 \eeq
 Evaluating this at the TRIM points we note the following interesting observations. The first is that $Q(0,\pi)=Q(0,\pi)$ and so the denominator in Eq. (\ref{chern1}) does not contribute to $C_1$.  Second, at the point $(\pi,\pi)$ we can make a similar argument to that made in Ref. [\onlinecite{ghosh}]. In units of the lattice constant, if $\pi^2 \gg M, B, \Delta_0, \mu$  then we can focus only on the $k^4$ term in $Q(\pi,\pi)$. This term is simply $t^2-B^2$. Finally we have $Q(0,0)= \mu^2-M^2$ leading to our result for $C_1$
 \beq
 C_1 =  \frac{1}{i\pi} \ln\text{sgn}\left[(t^2-B^2)( \mu^2-M^2)\right]
 \eeq
 Since physically we expect $|B|<t$ then the topology of the system is entirely determined by whether or not $\mu$ falls in the gap in the spin-orbit split bands opened by $M$. If $|\mu|<|M|$ (Fermi surface in the gap) then $C_1=1$ and the system has non-trivial topology. Note that $\Delta_s$ is not technically defined at $(0,0)$, this is likely an artifact of our continuum theory and we have replaced it with its limiting value here. We have checked our observations here using the numerical formula for the calculation of $C_1$ given in Eq. (11) of Ref. [\onlinecite{Farrell}].

 The parameter range $|M|>|\mu|$ is of course the type of system we have considered in our renormalization group approach in the previous section. Thus any superconductivity that develops in the system for this range of parameters will have non-trivial topology.

 \section{Conclusion}\label{sec:conclusion}
 We have studied a model of interacting, spin-orbit coupled electrons using renormalization group methods. After simplifying the model of Ref. [\onlinecite{Farrell}] by taking the continuum limit, we have applied the methods developed in Refs. [\onlinecite{wang1},\onlinecite{wang2},\onlinecite{raghu1},\onlinecite{raghu2}] to our model and focused on a system with a single Fermi surface. Our analysis shows that for this range of parameters the most dominant angular momentum pairing channel under RG has $p$-wave or $f$-wave symmetry depending on the sign of the Zeeman parameter $M$. Such an interaction should lead to a superconductor with intraband Cooper pairs and a $p$-wave or $f$-wave order parameter.

 To verify the topology of the state we simplified our model to include only the dominant interaction channel and performed a mean field analysis of the resultant effective Hamiltonian. This analysis shows that the system develops topological superconductivity. The condition for non-trivial topology in the physical case, $|B|<t$, is that $|\mu|<|M|$, that is that the Fermi surface must lie in the gap in the spin-orbit split bands opened by $M$. This is similar to the condition discussed by Sau\cite{Sau} and Alicea\cite{Alicea} in the context of spin-orbit coupled bands in proximity to a superconductor. Thus the results here provide added justification to the case that interactions, rather than proximity effect, may be used to obtain topological superconductivity\cite{Farrell,FarrellSC,tewari}. Moreover, we see the potential for the following physical scenario. In a topologically trivial superconductor with spin-orbit coupling the topology may change to a non-trivial one upon applying a Zeeman field.  This occurs since the Zeeman field provides the necessary gapping of one of the bands, leaving one band whose electron spins are locked to the momentum direction.


 \section{Acknowledgements}
 The authors are thankful for useful discussions with J.~Alicea, B.~A.~Bernevig and Y.~Bresler. Financial support for this work was provided by the NSERC and FQRNT (TPB) the Vanier Canada Graduate Scholarship (AF) and the Walter C.~Sumner Memorial Fellowship (AF). Numerical calculations for this work were performed using CLUMEQ/McGill HPC supercomputing resources.

\bibliographystyle{apsrev}
\bibliography{topoSC}

\begin{widetext}
\appendix
\section{Interaction Vertex in the Band Basis}

We now write the interaction Hamiltonian in terms of the new band operators. We note that our unitary transformation can be written as
 \beq
 c_{\kv,\alpha} = \sum_{\lambda} W_{\alpha,\lambda}(\kv) b_{\kv,\lambda}
 \eeq
 where $W_{\alpha,\lambda}(\kv) = \exp\left(\left(\frac{1-\sigma^z_{\alpha,\alpha}}{2} \right)\theta_{\kv} \right)f_{\sigma^z_{\alpha,\alpha}\lambda} (\kv)\eta_{\sigma, \lambda}$ with $\eta_{\downarrow, -}=-1$ and $\eta_{\alpha,\lambda}$ for all other combinations of $\alpha$ and $\lambda$. Meanwhile we have
  \beq
 c_{\kv,\alpha}^\dagger  = \sum_{\lambda} W^*_{\alpha,\lambda}(\kv) b^\dagger_{\kv,\lambda}
 \eeq
 Making use of the above we then have the interaction contribution

  \beqa
H_{int} &=&\sum_{\kv_1, \kv_2, \kv_3, \kv_4}\sum_{\lambda_1,\lambda_2,\lambda_3,\lambda_4}\delta_{\kv_1+\kv_2,\kv_3+\kv_4} \\ \nonumber &\times&{V}_{\lambda_1,\lambda_2,\lambda_3,\lambda_4}(\kv_1, \kv_2, \kv_3, \kv_4)   b_{\kv_1, \lambda_1}^\dagger b_{\kv_2, \lambda_2}^\dagger  b_{\kv_3, \lambda_3} b_{\kv_4, \lambda_4}
\eeqa
where
\begin{equation}
{V}_{\lambda_1,\lambda_2,\lambda_3,\lambda_4}(\kv_1, \kv_2, \kv_3, \kv_4)=\sum_{\alpha_1,\alpha_2,\alpha_3,\alpha_4}\;\;\;\; {U}_{\alpha_1,\alpha_2,\alpha_3,\alpha_4}(\kv_1, \kv_2, \kv_3, \kv_4) \;\;\;
W^*_{\alpha_1,\lambda_1}(\kv_1)  W^*_{\alpha_2,\lambda_2}(\kv_2) W_{\alpha_3,\lambda_3}(\kv_3)W_{\alpha_4,\lambda_4}(\kv_4).
\end{equation}
The above describes scattering events between electrons in the two bands; electrons in eigenstates $(\kv_4, \lambda_4)$ and $(\kv_3, \lambda_3)$ scatter to states  $(\kv_2, \lambda_2)$ and $(\kv_1, \lambda_1)$ with some associated interaction strength ${V}_{\lambda_1,\lambda_2,\lambda_3,\lambda_4}(\kv_1, \kv_2, \kv_3, \kv_4)  $ which depends on momenta and band index. Finally the delta function conserves momentum in this scattering process.
A more convenient form for the interaction strength $V$ is given by
\beqa
 \tilde{V}_{\lambda_1,\lambda_2,\lambda_3,\lambda_4}(\kv_1, \kv_2, \kv_3, \kv_4)&=&  \frac{U}{2N}  \sum_{\alpha_1}W^*_{\alpha_1,\lambda_1}(\kv_1)  W^*_{\bar{\alpha}_1,\lambda_2}(\kv_2)  \nonumber \\ &\times& W_{\bar{\alpha}_1,\lambda_3}(\kv_3)W_{\alpha_1,\lambda_4}(\kv_4)
 \eeqa
where we have used a bar symbol, $\bar{\alpha}$ to indicate the compliment to spin $\alpha$. Writing out the sum over $\alpha$ explicitly and defining  $F_{i,j} =  f_{\lambda_i} (\kv_i) f_{-\lambda_j} (\kv_j) $ we then obtain
  \beqa
&& {V}_{\lambda_1,\lambda_2,\lambda_3,\lambda_4}(\kv_1, \kv_2, \kv_3, \kv_4)=\\ \nonumber &&
     \frac{U \left(  \lambda_2e^{-\theta_{\kv_2} }F_{1,2} -\lambda_1e^{-\theta_{\kv_1} }  F_{2,1}  \right)\left(  \lambda_3e^{\theta_{\kv_3} } F_{4,3} -\lambda_4e^{\theta_{\kv_4} }  F_{3,4}  \right) }{4N}
 \eeqa
We see that $V$ is antisymmetric under the exchange of {\em either} indices 1 and 2 or 3 and 4 and symmetric under the exchange of both.

\section{Three-Loop Expansion for $S'_{int}$ on the Lattice}
\subsection{Tree Level and One-Loop}
We now give expressions for the first few terms in the cumulant expansion for $S'_{int}$. Here we give general expressions in terms of unsolved integrals and later on we make some simplifying specializations in order to perform these integrals approximately. As we are interested in superconductivity in this model we will be invested in how the BCS channel of the original interacting action evolves under renormalization. This channel is specified by $\kv_4=-\kv_3=\kv$, $\lambda_3=\lambda_4=\lambda$, $\kv_1=-\kv_2=\kv'$ and $\lambda_2=\lambda_1=\lambda'$ in the original (bare) interaction $S_{int}$. To this end we will set all external momenta accordingly.
 \begin{figure}[tb]
  \setlength{\unitlength}{1mm}
\vspace{5mm}
   \includegraphics[scale=.4]{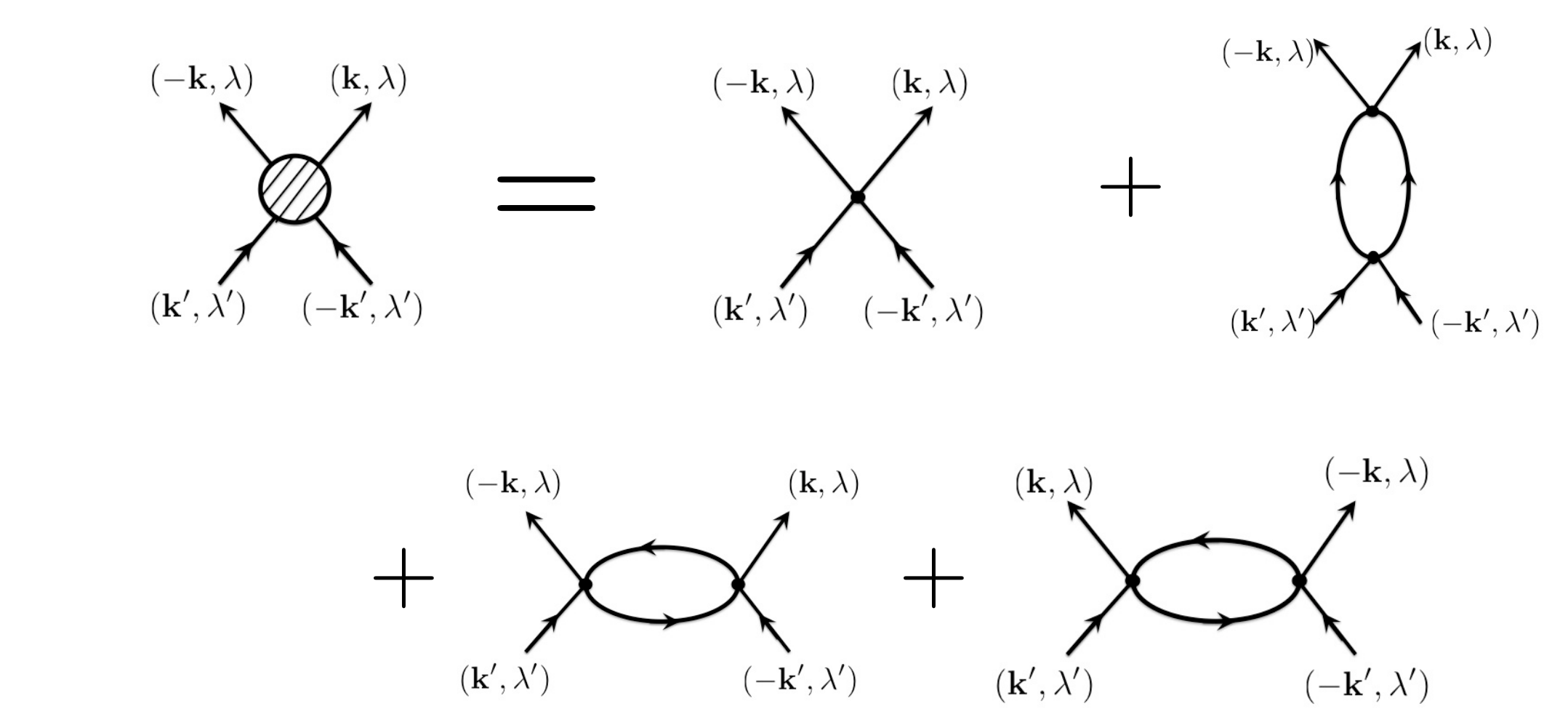}
\caption{{\small
Diagrams contributing to the renormalization of the BCS coupling function up to one-loop.
     }
     }\label{fig:oneloop}
\end{figure}

Let us begin our discussion with the effective interaction up to one-loop, afterwards we will extend this to three-loops. The diagrams contributing to this effective action are shown in Fig.~\ref{fig:oneloop}. We have
\beqa
S'_{int, 1\ell}& =&\delta S^{\text{tree}}_{int}+ \delta S^{\text{one-loop}}_{int} \\ \nonumber &=&   \int_0^\beta d\tau  \sum_{\kv, \kv', \lambda, \lambda'}(V^{\text{tree}}_{\lambda, \lambda'}(\kv,\kv')+ V^{\text{one-loop}}_{\lambda, \lambda'}(\kv,\kv'))\\ \nonumber &\times&b_{\kv, \lambda, <}^*(\tau)b_{-\kv, \lambda,<}^*(\tau)b_{-\kv', \lambda',<}(\tau)b_{\kv', \lambda', <}(\tau)
\eeqa
where  the tree level term gives the contribution $V^{\text{tree}}_{\lambda, \lambda'}(\kv,\kv')=\frac{U}{N} e^{i(\theta_{\kv'}-\theta_{\kv}) }\lambda \lambda' F_{\lambda,\lambda}(\kv,\kv)F_{\lambda', \lambda'}(\kv',\kv')  $ and the one-loop contribution is
\beqa
V^{\text{one-loop}}_{\lambda, \lambda'}(\kv,\kv') &=& \frac{U^2}{64N}\left(\Pi(-\kv,\kv')-\Pi(\kv,\kv')\right)\nonumber  \\ &-&\frac{U}{2}V^{\text{tree}}_{\lambda, \lambda'}(\kv,\kv')P(\Lambda, s)
\eeqa
In the above we have defined the following integral
\beqa
  \Pi_{\lambda, \lambda'}(\kv,\kv') &=&  \sum_{\lambda_5, \lambda_6}\int_{>} \frac{d^2\pv}{(2\pi)^2}\left(   \frac{n_F(E_{\pv, \lambda_5})-n_F(E_{\pv+\kv+\kv', \lambda_6})}{E_{\pv, \lambda_5}-E_{\pv+\kv+\kv', \lambda_6}} \right)\nonumber  \\ &\times&G_{\lambda, \lambda', \lambda_5, \lambda_6} (\kv,\kv',\pv)
\eeqa
with the function
\beqa
G_{\lambda, \lambda', \lambda_5, \lambda_6} (\kv,\kv',\pv) &=& 16w_{\lambda_5, \lambda}(\pv, \kv)w^*_{\lambda', \lambda_6}(-\kv', \kv+\kv'+\pv) \nonumber  \\&\times&w_{\lambda_6, \lambda}(\kv+\kv'+\pv, -\kv)w^*_{\lambda', \lambda_5}(\kv', \pv)   \nonumber \\
\eeqa
and the $>$ on the $\pv$ integral is to remind us that the integral must be performed over the regions of $\pv$-space satisfying both $\Lambda/s\le |E_{\pv, \lambda_5} |\le \Lambda$ and $\Lambda/s\le E_{\pv+\kv+\kv', \lambda_6} \le \Lambda$. We have also defined the momentum independent integral $P$
\beqa
P(\Lambda, s) &=&  \sum_{\lambda_5, \lambda_6 }\int _{>} \frac{d^2 \pv}{(2\pi)^2} \left(\frac{1-n_F(E_{\pv, \lambda_5})-n_F(E_{\pv, \lambda_6})}{E_{\pv, \lambda_5}+E_{\pv, \lambda_6}}\right)\nonumber\\  &\times&|w_{\lambda_5, \lambda_6}(-\pv,\pv)|^2
\eeqa

\subsection{Two-Loop}
We now move on to higher our diagrams of which we keep only diagrams with logarithmic divergences. The diagrams we sum are essentially the same as the diagrams used in Refs. [\onlinecite{wang1,wang2}] and we do not redraw them here.

As a two-loop contribution to our interaction we obtain
  \beqa
\delta S^{\text{two-loop}}_{int} &=&   \int_0^\beta d\tau  \sum_{\kv, \kv', \lambda, \lambda'} V^{\text{two-loop}}_{\lambda, \lambda'}(\kv,\kv') \\ &\times&\nonumber b_{\kv, \lambda, <}^*(\tau)b_{-\kv, \lambda,<}^*(\tau)b_{-\kv', \lambda',<}(\tau)b_{\kv', \lambda', <}(\tau)
\eeqa
where we have defined
\beqa
V^{\text{two-loop}}_{\lambda, \lambda'}(\kv,\kv') &=& \frac{U^3}{4N}\lambda\lambda' e^{i(\theta_{\kv'}-\theta_{\kv})} F_{\lambda\lambda}(\kv,\kv)F_{\lambda'\lambda'}(\kv',\kv')P^2(\Lambda,s) \nonumber \\ &-&\frac{U^3}{128N}(\lambda' e^{i\theta_{\kv'}} F_{\lambda'\lambda'}(\kv',\kv') I^{2\ell}(\kv,\lambda)\\ \nonumber &+&\lambda e^{-i\theta_{\kv}} F_{\lambda\lambda}(\kv,\kv) (I^{2\ell}(\kv',\lambda'))^*)
\eeqa
where we have the integral
\beqa
{I}^{2\ell} (\kv,\lambda) &=&\sum_{\mu_1, \mu_2} \int_> \frac{d^2\pv}{(2\pi)^2} \left(\frac{1-n_f(E_{\pv,\mu_2})-n_f(E_{\pv,\mu_1})}{E_{\pv,\mu_2}+E_{\pv,\mu_1}}\right)w_{\mu_2, \mu_1}(\pv,-\pv)( \hat{\Pi}_{\lambda,\mu_1,\mu2}(-\kv,\pv)-\hat{\Pi}_{\lambda,\mu_1,\mu2}(\kv,\pv))\nonumber  \\
\hat{\Pi}_{\lambda,\mu_1,\mu2}(\kv,\pv)  &=& \sum_{\mu_3, \mu_4} \int_> \frac{d^2\pv_2}{(2\pi)^2} \left(\frac{n_f(E_{\pv_2,\mu_3})-n_f(E_{\pv_2+\pv+\kv,\mu_4})}{E_{\pv_2,\mu_3}-E_{\pv_2+\pv+\kv,\mu_4}}\right)\hat{G}_{\lambda,\mu_1...\mu_4}(\kv,\pv,\pv_2)\\ \nonumber
\hat{G}_{\lambda,\mu_1...\mu_4}(\kv,\pv,\pv_2) &=& G_{\lambda, \mu_1,\mu_3,\mu_4}(\kv,\pv,\pv_2)\delta_{\mu_1,\mu_2} + \tilde{G}_{\lambda, \mu_1,\mu_3,\mu_4}(\kv,\pv,\pv_2)\sigma^{x}_{\mu_1,\mu2}
\eeqa
where $G_{\lambda, \mu_1,\mu_3,\mu_4}(\kv,\pv,\pv_2)$ is defined above and

\beqa
\tilde{G}_{\lambda, \lambda', \lambda_5, \lambda_6} (\kv,\kv',\pv) &=& 16w_{\lambda_5, \lambda}(\pv, \kv) \\ \nonumber &\times&w^*_{-\lambda', \lambda_6}(-\kv', \kv+\kv'+\pv) \nonumber  \\&\times&w_{\lambda_6, \lambda}(\kv+\kv'+\pv, -\kv)w^*_{\lambda', \lambda_5}(\kv', \pv)  \nonumber
\eeqa

\subsection{Three-Loop}
We now finally move on to fourth order. Their contribution is as follows
  \beqa
\delta S^{\text{three-loop}}_{int} &=&   \int_0^\beta d\tau  \sum_{\kv, \kv', \lambda, \lambda'} V^{\text{three-loop}}_{\lambda, \lambda'}(\kv,\kv')\nonumber \\ &\times&b_{\kv, \lambda, <}^*(\tau)b_{-\kv, \lambda,<}^*(\tau)b_{-\kv', \lambda',<}(\tau)b_{\kv', \lambda', <}(\tau)
\eeqa
where we have defined
\beqa
V^{\text{three-loop}}_{\lambda, \lambda'}(\kv,\kv') &=&-\frac{U^4\lambda\lambda' e^{i(\theta_{\kv'}-\theta_{\kv})} F_{\lambda\lambda}(\kv,\kv)F_{\lambda'\lambda'}(\kv',\kv')}{2^3N} \left(P^3(\Lambda,s)+ \frac{ \hat{P}(\Lambda,s)}{8}  \right)  \nonumber \\ &+&\frac{U^4P(\Lambda,s) }{2^8N}(2 \lambda' e^{i\theta_{\kv'}} F_{\lambda'\lambda'}(\kv',\kv')I^{2\ell}(\kv,\lambda)+2 \lambda e^{-i\theta_{\kv}} F_{\lambda\lambda}(\kv,\kv)( I^{2\ell}(\kv',\lambda'))^*) \\ \nonumber &+& \frac{U^4}{2^{11} N} I^{3\ell}(\kv,\kv',\lambda,\lambda')
\eeqa
where we have the new integrals
\beqa
I^{3\ell}(\kv,\kv',\lambda,\lambda') &=& \sum_{\mu_1, \mu_2}  \int_> \frac{d^2\pv}{(2\pi)^2}  \left(\frac{1-n_f(E_{\pv,\mu_2})-n_f(E_{\pv,\mu_1})}{E_{\pv,\mu_2}+E_{\pv,\mu_1}}\right)  \\ \nonumber &\times&(\hat{\Pi}_{\lambda,\mu_1,\mu_2}(\kv,\pv)-\hat{\Pi}_{\lambda,\mu_1,\mu_2}(-\kv,\pv))(\hat{\Pi}^*_{\lambda', \mu_2,\mu_1}(\kv', -\pv)-\hat{\Pi}^*_{\lambda', \mu_2,\mu_1}(-\kv', -\pv))\nonumber \\ \nonumber
\hat{P}(\Lambda, S) &=& \sum_{\mu_1...\mu_6}  \int_> \frac{d^2\pv_1}{(2\pi)^2} \int_> \frac{d^2\pv_2}{(2\pi)^2} \int_> \frac{d^2\pv_3}{(2\pi)^2}w_{\mu_2, \mu_1}(\pv_1, -\pv_1) w^*_{\mu_5, \mu_6}(\pv_3, -\pv_3)\\ \nonumber &\times&
 \left(\frac{1-n_f(E_{\pv_1,\mu_2})-n_f(E_{\pv_1,\mu_1})}{E_{\pv_1,\mu_2}+E_{\pv_1,\mu_1}}\right)\left(\frac{1-n_f(E_{\pv_3,\mu_6})-n_f(E_{\pv_3,\mu_5})}{E_{\pv_3,\mu_6}+E_{\pv_3,\mu_5}}\right) \\ \nonumber &\times&
\left(\frac{n_f(E_{\pv_2,\mu_3})-n_f(E_{\pv_2+\pv_1+\pv_3,\mu_4})}{E_{\pv_2,\mu_3}-E_{\pv_2+\pv_1+\pv_3,\mu_4}}\right)\hat{\mathcal{G}}_{\mu_1...\mu_6}(\pv_1,\pv_2, \pv_3)
\\
\hat{\mathcal{G}}_{\mu_1...\mu_6}(\pv_1,\pv_2, \pv_3) &=&\delta_{\mu_5, \mu_6} \hat{G}_{\mu_5,\mu_1\mu_2\mu_3\mu_4}(\pv_3,\pv_1,\pv_2) +\delta_{\mu_1, \mu_2}\sigma^x_{\mu_5,\mu_6} \tilde{G}^*_{\mu_1,\mu_5,\mu_3,\mu_4}(\pv_1,\pv_3,\pv_2)\delta_{\mu_1,\mu_2}\\ \nonumber & +&\sigma^x_{\sigma_5,\sigma_6}\sigma^x_{\mu_1,\mu_2} \bar{G}_{\mu_5 \mu_1,\mu_3,\mu_4}(\pv_3,\pv_1,\pv_2)
\eeqa
where

\beqa
\bar{G}_{\lambda, \lambda', \lambda_5, \lambda_6} (\kv,\kv',\pv) &=& 16w_{\lambda_5, \lambda}(\pv, \kv) \\ \nonumber &\times&w^*_{-\lambda', \lambda_6}(-\kv', \kv+\kv'+\pv) \nonumber  \\&\times&w_{\lambda_6, -\lambda}(\kv+\kv'+\pv, -\kv)w^*_{\lambda', \lambda_5}(\kv', \pv)  \nonumber
\eeqa

Using the above expressions for the diagrams up to fourth order in $U$ and logarithmically enhanced we have the effective BCS channel coupling
\beqa\label{Vlattice}
V'_{\lambda,\lambda'}(\kv,\kv', \Lambda, s) &=&\frac{U}{N} e^{i(\theta_{\kv'}-\theta_{\kv}) }\lambda \lambda' F_{\lambda,\lambda}(\kv,\kv)F_{\lambda', \lambda'}(\kv',\kv')\left(1 - \frac{U}{2}P(\Lambda, s)+\frac{U^2}{4}P^2(\Lambda,s) - \frac{U^3}{8}P^3(\Lambda,s)- \frac{U^3}{64} \hat{P}(\Lambda,s)\right)\nonumber  \\ \nonumber &+& \frac{U^2}{64N}I^{1\ell}(\kv,\kv') + \frac{U^4}{2^{11} N} I^{3\ell}(\kv,\kv',\lambda,\lambda')   \nonumber \\ &-&\frac{U^3}{128N}\left(\lambda' e^{i\theta_{\kv'}}F_{\lambda'\lambda'}(\kv',\kv') I^{2\ell}(\kv,\lambda)+ \lambda e^{-i\theta_{\kv}}F_{\lambda\lambda}(\kv,\kv)( I^{2\ell}(\kv',\lambda'))^*\right)\left(1-UP(\Lambda,s)  \right)
\eeqa
where $I^{1\ell}(\kv,\kv')= \left(\Pi(-\kv,\kv')-\Pi(\kv,\kv')\right)$.

\section{Continuum Limit}
\subsection{Dispersions}
The integrals involved in the discussion above are formidable and do not allow any further analytic progress. To make progress we focus on the continuum limit of the model above by sending $\sin{k_i}\to k_i$, $\cos{k_i} \to 1-k_i^2/2$. This gives the dispersion
\beq
E_{ \kv, \lambda} = tk^2 -\tilde{\mu}-4t +\lambda \sqrt{(A^2-2BM)k^2+M^2+B^2k^4}
\eeq
where $k=\sqrt{k_x^2+k_y^2}$. To make connection with standard conventions we redefine parameters as follows and ${\mu} =\tilde{ \mu}+4t$. Then we have
\beqa
E_{ \kv, \lambda} &=& t k^2-{\mu} +\lambda \sqrt{(A^2-2BM)k^2+M^2+B^2k^4} \nonumber \\  &=& \epsilon_{\kv, \lambda} -{\mu}= E_{\lambda}(k)
\eeqa
where $ \epsilon_{\kv, \lambda} = t k^2+\lambda \sqrt{(A^2-2BM)k^2+M^2+B^2k^4}$. In this limit we have the new definitions
\beqa
e^{i\theta_{\kv}} &=& \frac{k_x+ik_y}{\sqrt{k_x^2+k_y^2}} \\ \nonumber
f_{\lambda}(\kv) &=& f_{\lambda}(k)=\sqrt{\frac{d+\lambda d_3}{2d}}
\eeqa
where $d= \sqrt{(A^2-2BM)k^2+M^2+B^2k^4}$. At $B=0$ this is precisely the band structure considered in the work by Sau {\it et al}.

\subsection{Rotational Invariance}
As discussed in the main text in the continuum limit we obtain a theory which depends only on the angle between given wave vectors on the Fermi surface. In developing this result it is incredibly useful to realize that
\beqa
I^{1\ell}_{\lambda,\lambda'}(\kv, \kv',s) &=& e^{i\phi} I^{1\ell}_{\lambda, \lambda'}(k,k',\phi,s) \\ \nonumber
I^{2\ell}_{\lambda}(\kv,s) &=& e^{-i\theta_{\kv}} I^{2\ell}_{\lambda}(k,s) \\ \nonumber
I^{3\ell}_{\lambda,\lambda'}(\kv, \kv',s) &=&-e^{i\phi} I^{3\ell}_{\lambda, \lambda'}(k,k',\phi,s)
\eeqa
where $\phi=\theta_{\kv}-\theta_{\kv'}$. The manipulations required to show this is identical to that outlined in [\onlinecite{wang2}]. We will give results here for $I^{1\ell}(\kv,\kv',s)$ first. We begin by noting the following result
\beqa
  \Pi_{\lambda, \lambda'}(-\kv,\kv')  &=&   \sum_{\lambda_5, \lambda_6}\int_{>} \frac{d^2\pv}{(2\pi)^2}\left(   \frac{n_F(E_{ \lambda_5}(p))-n_F(E_{ \lambda_6}(\sqrt{p^2+Q^2-2pQ\cos{(\theta_{\pv}-\theta_{\Qv})}}))}{E_{ \lambda_5}(p)-E_{ \lambda_6}(\sqrt{p^2+Q^2-2pQ\cos{(\theta_{\pv}-\theta_{\Qv})}})} \right)\nonumber  16w_{\lambda_5, \lambda}(\pv, -\kv)w^*_{\lambda', \lambda_6}(-\kv', \pv-\Qv) \nonumber  \\&\times&w_{\lambda_6, \lambda}(\pv-\Qv, \kv)w^*_{\lambda', \lambda_5}(\kv', \pv) \\ \nonumber
\eeqa
where $\Qv=\kv-\kv'$. By making use of the identity
\beq
e^{i\theta_{\kv_1-\kv_2}} = \frac{k_1 e^{i\theta_{\kv_1}}-k_2e^{i\theta_{\kv_2}}}{\sqrt{k_1^2+k_2^2-2k_1k_2\cos(\theta_{\kv_1}-\theta_{\kv_2})}},
\eeq
shifting $\theta_{\pv} \to \theta_{\pv}+\theta_{\Qv}$ and then using
 \beq
e^{i\theta_{\Qv}} =\left(\frac{k e^{i\theta_{\kv}}-k'e^{i\theta_{\kv'}}}{Q}\right)
\eeq
one can show using some straightforward but tedious manipulations that if we define
\beqa
\mathcal{F}_{\mu_1...\mu_6}(k,k', \pv, \phi) &=& \frac{16}{q^2} \left(\mu_5 e^{-i\theta_{\pv}}g F_{\mu_1, \mu_5}(k,p)+\mu_1 e^{i\phi} F_{\mu_5, \mu_1}(p,k) \right)\left(
   \mu_3F_{\mu_5,\mu_3}(p,k') - \mu_5 e^{i\theta_{\pv}} g^*F_{\mu_3,\mu_5}(k',p)\right) \\ \nonumber &\times&
\left(  -\mu_4 q F_{\mu_6,\mu_4}( q,k') + \mu_6g^*{Q}F_{\mu_4,\mu_6}(k', q) - \mu_6g^* e^{i\theta_{\pv}} pF_{\mu_4,\mu_6}(k', q)\right)\\ \nonumber &\times& \left(
   \mu_6e^{-i\phi} g e^{-i\theta_{\pv}}pF_{\mu_2,\mu_6}(k, q)- \mu_6e^{-i\phi} gQF_{\mu_2,\mu_6}(k, q) -\mu_2  q F_{\mu_6,\mu_2}( q,k)\right)
\eeqa
where $q=\sqrt{p^2+Q^2-2Qp\cos(\theta_{\pv})}$, $\phi=\theta_{\kv'}-\theta_{\kv}$, $Q=\sqrt{k^2+k'^2-2k'\cos(\phi)}$ and $g=\frac{k e^{i\phi}-k'}{Q}$ then it follows that
 \beqa
&&  \Pi_{\lambda, \lambda'}(-\kv,\kv')  = e^{i\phi }  \sum_{\lambda_5, \lambda_6}\int_{>} \frac{d^2\pv}{(2\pi)^2}\left(   \frac{n_F(E_{ \lambda_5}(p))-n_F(E_{ \lambda_6}(q))}{E_{ \lambda_5}(p)-E_{ \lambda_6}(q)} \right)\nonumber  \\ &\times&\mathcal{F}_{\lambda,\lambda,\lambda',\lambda',\lambda_5,\lambda_6}(k,k', \pv, \phi)=e^{i\phi } V_{\lambda,\lambda,\lambda',\lambda'}(k,k', \phi)
\eeqa

Following the exact same analysis one can show that
 \beqa
 && \hat{\Pi}_{\lambda, \mu_1,\mu_2}(-\kv,\kv')  =e^{i\phi }  \sum_{\lambda_5, \lambda_6}\int_{>} \frac{d^2\pv}{(2\pi)^2}\left(   \frac{n_F(E_{ \lambda_5}(p))-n_F(E_{ \lambda_6}(q))}{E_{ \lambda_5}(p)-E_{ \lambda_6}(q)} \right)\mathcal{F}_{\lambda,\lambda,\mu_1,\mu_2,\lambda_5,\lambda_6}(k,k', \pv, \phi)= e^{i\phi } V_{\lambda,\lambda,\mu_1,\mu_2}(k,k', \phi)\\   \nonumber
&&   \tilde{\Pi}_{ \mu_1,\mu_2, \mu_5,\mu_6}(-\kv,\kv')  = e^{i\phi }  \sum_{\lambda_5, \lambda_6}\int_{>} \frac{d^2\pv}{(2\pi)^2}\left(   \frac{n_F(E_{ \lambda_5}(p))-n_F(E_{ \lambda_6}(q))}{E_{ \lambda_5}(p)-E_{ \lambda_6}(q)} \right)\mathcal{F}_{\mu_5,\mu_6,\mu_1,\mu_2,\lambda_5,\lambda_6}(k',k, \pv, \phi)= e^{i\phi } V_{\mu_5,\mu_6,\mu_1,\mu_2}(k',k, \phi)
\eeqa
These three results have immediate implications for the $I^{n\ell}$ terms. First $I^{1\ell}(\kv,\kv', s) = e^{i\phi} I^{1\ell}(k,k',\phi, s) $ where $I^{1\ell}(k,k',\phi, s)=V_{\lambda,\lambda'}(k,k', \phi, s)+V_{\lambda,\lambda'}(k,k', \phi+\pi,s )$ and $V_{\lambda,\lambda'}(k,k', \phi) =  V_{\lambda,\lambda,\lambda',\lambda'}(k,k', \phi)$. Next we have for $I^{2\ell}$
\beqa
{I}^{2\ell} (\kv,\lambda, s) &=&e^{-i\theta_{\kv}}\sum_{\mu_1, \mu_2} \int_> \frac{d^2\pv}{(2\pi)^2} \left(\frac{1-n_f(E_{\pv,\mu_2})-n_f(E_{\pv,\mu_1})}{E_{\pv,\mu_2}+E_{\pv,\mu_1}}\right)|w_{\mu_2, \mu_1}(\pv,-\pv)|\\ \nonumber &\times& (  V_{\lambda,\lambda,\mu_1,\mu_2}(k,p, \theta_{\pv}-\theta_{\kv})+  V_{\lambda,\lambda,\mu_1,\mu_2}(k,p, \theta_{\pv}-\theta_{\kv}+\pi))
\eeqa
shifting $\theta_{\pv}\to \theta_{\pv}+\theta_{\kv}$ above then gives ${I}^{2\ell} (\kv,\lambda, s) = e^{-i\theta_{\kv}} I_{\lambda}^{2\ell}(k,s)$. Finally we have
\beqa
I^{3\ell}(\kv,\kv',\lambda,\lambda', s) &=&- e^{i(\theta_{\kv'}-\theta_{\kv})}\sum_{\mu_1, \mu_2}  \int_> \frac{d^2\pv}{(2\pi)^2}  \left(\frac{1-n_f(E_{\pv,\mu_2})-n_f(E_{\pv,\mu_1})}{E_{\pv,\mu_2}+E_{\pv,\mu_1}}\right)  \\ \nonumber &\times&(  V_{\lambda,\lambda,\mu_1,\mu_2}(k,p, \theta_{\pv}-\theta_{\kv})+  V_{\lambda,\lambda,\mu_1,\mu_2}(k,p, \theta_{\pv}-\theta_{\kv}+\pi))\\ \nonumber &\times&(  V_{\lambda,\lambda,\mu_1,\mu_2}(k',p, \theta_{\pv}-\theta_{\kv'})+  V_{\lambda,\lambda,\mu_1,\mu_2}(k',p, \theta_{\pv}-\theta_{\kv'}+\pi))^*\nonumber \\ \nonumber
\eeqa
Shifting  $\theta_{\pv}\to \theta_{\pv}+\theta_{\kv'}$ above then immediately gives $I^{3\ell}(\kv,\kv',\lambda,\lambda', s) =- e^{i(\theta_{\kv'}-\theta_{\kv})}I^{3\ell}_{\lambda,\lambda'}(k,k', \phi, s)$.

By using the above results we can obtain a simplified continuum limit version of Eq. (\ref{Vlattice}) which is given as follows
\beqa
V'_{\lambda,\lambda'}(k,k',\phi, \Lambda, s) &=&\frac{U}{N} e^{i\phi }\lambda \lambda' F_{\lambda,\lambda}(k,k)F_{\lambda', \lambda'}(k',k')\left(1 - \frac{U}{2}P(\Lambda, s)+\frac{U^2}{4}P^2(\Lambda,s) - \frac{U^3}{8}P^3(\Lambda,s)- \frac{U^3}{64} \hat{P}(\Lambda,s)\right)\nonumber  \nonumber \\ &-&\frac{U^3e^{i\phi}}{128N}\left(\lambda' F_{\lambda'\lambda'}(k',k') I^{2\ell}_{\lambda}(k)+ \lambda F_{\lambda\lambda}(k,k)( I^{2\ell}_{\lambda'}(k'))^*\right)\left(1-UP(\Lambda,s)  \right)  \\ \nonumber &+& \frac{U^2}{64N}e^{i\phi} I^{1\ell}_{\lambda, \lambda'}(k,k',\phi) - \frac{U^4}{2^{11} N}e^{i\phi} I^{3\ell}_{\lambda, \lambda'}(k,k', \phi)
\eeqa
We see that $V'$ depends only on the relative angle $\phi$, as is to be expected. We proceed to write $V'_{\lambda,\lambda'}(k,k',\phi, \Lambda, s) = \frac{e^{i\phi}U}{N} v'_{\lambda,\lambda'}(k,k',\phi, \Lambda, s)$.

\subsection{Evaluation of Integrals}
We now work to evaluate, or approximately evaluate, the integrals above. The first integral we focus on is $P(\Lambda, s)$. We have
\beqa
P(\Lambda, s) &=&  \sum_{\lambda_5, \lambda_6 }\int _{>} \frac{d^2 \pv}{(2\pi)^2} \left(\frac{1-n_F(E_{\pv, \lambda_5})-n_F(E_{\pv, \lambda_6})}{E_{\pv, \lambda_5}+E_{\pv, \lambda_6}}\right)\nonumber\\  &\times&|w_{\lambda_5, \lambda_6}(-\pv,\pv)|^2
\eeqa
where the `$>$' means {\em both} $E_{\pv, \lambda_5}$ and $E_{\pv, \lambda_6}$ must lie in the fast mode range $\Lambda/s<|E_{\lambda}(p)|<\Lambda$. We set $\lambda_6=\lambda_5$ in the above for the following reason. The $\pv$ integral is limited to values in $\pv$-space where both bands lie in the fast mode range. For two different bands, {\it i.e.} $\lambda_5\ne\lambda_6$ the small window of $\pv$ values for which $E_{\pv, \lambda_5}$ and $E_{\pv, \lambda_6}$ are in the fast range will in general be different. To simplify our calculation we assume that there is no overlap between these two regions and thus set
 $\lambda_6=\lambda_5$. The next simplification we make has to do with what part of the above integral we are interested in. We are interested in terms that diverge as we send $s\to\infty$\cite{raghu1}. The term  $|w_{\lambda_5, \lambda_5}(-\pv,\pv)|^2 $ is regular as $\pv$ approaches a Fermi wave vector and so in order to simplify matters we set $|w_{\lambda_5, \lambda_5}(-\pv,\pv)|^2  = |w_{\lambda_5, \lambda_6}(k_f^{\lambda_5}, k_f^{\lambda_5})|^2 =(2\lambda_5 F_{\lambda_5, \lambda_5})^2  =F_{\lambda_5}^2$ where we have defined $F_{\lambda_5}=2\lambda_5 F_{\lambda_5, \lambda_5}$. This leaves
 \beqa
P(\Lambda, s) &=&  \sum_{\lambda_5 } F_{\lambda_5}^2\int _{>} \frac{d^2 \pv}{(2\pi)^2} \left(\frac{1-2n_F(E_{\pv, \lambda_5})}{2E_{\pv, \lambda_5}}\right)
\eeqa
Now using the usual set of approximations for integrals of this type we set $\int _{>} \frac{d^2 \pv}{(2\pi)^2} \left(\frac{1-2n_F(E_{\pv, \lambda_5})}{2E_{\pv, \lambda_5}}\right) =N_{\lambda_5} \int _{>} dE\left(\frac{1-2n_F(E)}{2E}\right)= N_{\lambda_5} \ln(s)$ where $N_{\lambda_5}$ is the density of states at the Fermi surface. Thus
 \beqa
P(\Lambda, s) &=&  \sum_{\lambda_5 } F_{\lambda_5}^2N_{\lambda_5} \ln(s) \equiv p \ln(s)
\eeqa
where $p=\sum_{\lambda_5 } F_{\lambda_5}^2N_{\lambda_5}$. Next we work on simplifying ${I}^{2\ell} (k,\lambda, s) $. It is given by
\beqa
{I}^{2\ell}_{\lambda}  (k, s) &=&\sum_{\mu_1, \mu_2} \int_> \frac{d^2\pv}{(2\pi)^2} \left(\frac{1-n_f(E_{\pv,\mu_2})-n_f(E_{\pv,\mu_1})}{E_{\pv,\mu_2}+E_{\pv,\mu_1}}\right)|w_{\mu_2, \mu_1}(\pv,-\pv)|\\ \nonumber &\times& (  V_{\lambda,\lambda,\mu_1,\mu_2}(k,p, \theta_{\pv}, s)+  V_{\lambda,\lambda,\mu_1,\mu_2}(k,p, \theta_{\pv}+\pi,s))
\eeqa
Again we argue that because of the restriction on the states $E_{\pv,\mu_2}$ and $E_{\pv,\mu_1}$ to be fast modes we must set $\mu_1=\mu_2$. Further, only considering the part of the integral divergent as $s\to\infty$ while setting all other values of $\pv$ to lie on the Fermi surface gives
\beqa
{I}^{2\ell}_{\lambda} (k, s)&=&\ln(s) \sum_{\mu} F_{\mu}N_{\mu} \\ \nonumber &\times& \int_0^{2\pi}\frac{d\theta_\pv}{2\pi}  I^{1\ell}_{\lambda, \mu}(k, k_f^{\mu}, \theta_{\pv},s)
\eeqa
Taking the same steps with $I^{3\ell}_{\lambda,\lambda'}(k,k', \phi, s)$ gives us
\beqa
I^{3\ell}_{\lambda,\lambda'}(k,k', \phi, s) &=&\ln(s) \sum_{\mu}N_{\mu}  \int_{0}^{2\pi} \frac{d\theta_{\pv}}{2\pi}   I^{1\ell}_{\lambda, \mu}(k, k_f^{\mu}, \theta_{\pv}+\phi,s )( I^{1\ell}_{\lambda', \mu}(k', k_f^{\mu}, \theta_{\pv}, s))^*\eeqa
In order to simplify the above it becomes useful to Fourier transform $I^{1\ell}_{\lambda, \mu}(k, k', \theta )$ as follows
\beq
I^{1\ell}_{\lambda, \mu}(k, k', \theta, s ) = \sum_{m_z} e^{im_z \theta}v_{\lambda,\mu}^{m_z}(k, k', s)
\eeq
Then ${I}^{2\ell}_{\lambda} (k, s) = \ln(s) \sum_{\mu} F_{\mu}N_{\mu} v_{\lambda,\mu}^{0}(k, k',s)$ and also
\beqa
I^{3\ell}_{\lambda,\lambda'}(k,k', \phi, s) &=&\ln(s) \sum_{\mu, m_z} e^{im_z \phi} N_{\mu} v_{\lambda,\mu}^{m_z}(k, k_f^{\mu},s)( v_{\lambda',\mu}^{m_z}(k', k_f^{\mu},s))^*
\eeqa
Finally we work at simplifying $\hat{P}(\Lambda, s) $. Making our usual set of simplifying assumptions we have
\beqa
\hat{P}(\Lambda, s) &=&-e^{i(\theta_{\kv'}-\theta_{\kv})}\ln^2(s) \sum_{\mu,\nu} N_{\nu}F_{\nu}N_{\mu} F_{\mu}  \int_0^{2\pi} \frac{d\theta_{\pv_1}}{2\pi} \int_0^{2\pi} \frac{d\theta_{\pv_3}}{2\pi}V_{\mu,\nu}(k_f^{\mu},k_f^{\nu}, \theta_{\pv_3}-\theta_{\pv_1}-\pi,s)
\eeqa
If we redefine the angles $\Theta= (\theta_{\pv_1}+\theta_{\pv_3})/2$ and $\theta=\theta_{\pv_3}-\theta_{\pv_1}$ we immediately obtain
\beqa
\hat{P}(\Lambda, s) &=&-\frac{1}{2}e^{i(\theta_{\kv'}-\theta_{\kv})}\ln^2(s) \\ \nonumber &\times&\sum_{\mu,\nu} N_{\nu}F_{\nu}N_{\mu} F_{\mu} v^{0}_{\mu,\nu}(k_f^{\mu},k_f^{\nu},s)
\eeqa

Reflecting on the above we see that all of the integrals of interest in our 3-loop expansion of the effective interaction simplify to terms involving the single integral $v^{m_z}_{\mu,\nu}(k,k', s)$. The expression for this term is as follows
 \beqa
v^{m_z}_{\mu,\nu}(k,k', s)=  \sum_{\lambda_5, \lambda_6}\int_0^{2\pi} \frac{d\phi}{2\pi}e^{-im_z\phi}\int_{>} \frac{d^2\pv}{(2\pi)^2}\left(   \frac{n_F(E_{ \lambda_5}(p))-n_F(E_{ \lambda_6}(q))}{E_{ \lambda_5}(p)-E_{ \lambda_6}(q)} \right)\mathcal{F}_{\mu,\nu, \lambda_5,\lambda_6}(k,k', \pv, \phi)
\eeqa
where $\mathcal{F}_{\lambda,\lambda',\lambda_5,\lambda_6}(k,k', \pv, \phi)=\mathcal{F}_{\lambda,\lambda,\lambda',\lambda',\lambda_5,\lambda_6}(k,k', \pv, \phi)$. The above function is normal as we let $s\to\infty$ and so for our purposes it is sufficient to replace it with it's $s\to\infty$ counterpart\cite{raghu1, raghu2, wang1, wang2}. Further as we are only interested in incoming/outgoing momenta on the Fermi surface(s) we set $k$ and $k'$ appropriately. Defining $v^{m_z}_{\mu,\nu} =v^{m_z}_{\mu,\nu}(k_f^{\mu},k_f^{\nu}, s\to\infty) $we have
 \beqa
v^{m_z}_{\mu,\nu}=  \sum_{\lambda_5, \lambda_6}\int_0^{2\pi} \frac{d\phi}{2\pi}e^{-im_z\phi}\int \frac{d^2\pv}{(2\pi)^2}\left(   \frac{n_F(E_{ \lambda_5}(p))-n_F(E_{ \lambda_6}(q))}{E_{ \lambda_5}(p)-E_{ \lambda_6}(q)} \right)\mathcal{F}_{\mu,\nu, \lambda_5,\lambda_6}(k_f^{\mu},k_f^{\nu}, \pv, \phi)
\eeqa
where the restriction on the integral in $V$ has been dropped because as $s\to\infty$ all momenta satisfy $\Lambda/s<|E_{\lambda_5}(p)|<\Lambda$.

\subsection{Flow Equations}

By using the above results we can write the following
\beqa
v'_{\lambda,\lambda'}(\phi,  s) &=&\frac{U F_{\lambda} F_{\lambda'}}{4} \left(1 - \frac{Up}{2}\ln(s)+\frac{U^2p^2}{4}
\ln^2(s)- \frac{U^3p^3}{8}\ln^3(s)\right) + \frac{U^4F_{\lambda} F_{\lambda'}}{256} \sum_{\mu,\nu} N_{\mu} F_{\mu} v_{\mu, \nu}^{0} N_{\nu} F_{\nu} \nonumber  \nonumber \\ &-&
\frac{U^3}{256 }\ln(s) \left(F_{\lambda} \sum_{\mu} F_{\mu} N_{\mu} v_{\mu, \lambda'}^{0} +F_{\lambda'} \sum_{\mu}v_{\lambda,\mu}^{0} F_{\mu} N_{\mu}  \right)\left(2-Up\ln(s)  \right)  \\ \nonumber &+& \frac{U^2}{64} \sum_{m_z} e^{im_z\phi} v^{m_z}_{\lambda, \lambda'}- \ln(s)\frac{U^4}{2^{11} } \sum_{\mu, m_z}  e^{im_z\phi} N_{\mu}v_{\lambda,\mu}^{m_z} v_{\mu, \lambda'}^{m_z}
\eeqa
where in the above $m_z$ denotes an integer value, $k_f^{\lambda}$ is the Fermi wave vector magnitude for the $\lambda$ band, and we recall that we have defined $F_{\lambda} = 2\lambda F_{\lambda, \lambda}(k_f^{\lambda}, k_f^{\lambda})$, and  $p=\sum_{\mu} N_{\mu} F_{\mu}^2$.

We first consider the RG flow of the $m_z\ne0$ channel which is given by
\beqa
v'_{\lambda,\lambda'}( s, m_z) &=&\frac{U^2}{64}v^{m_z}_{\lambda, \lambda'}- \ln(s) \frac{U^4}{2^{11} } \sum_{\mu}  N_{\mu}v_{\lambda,\mu}^{m_z} v_{\mu, \lambda'}^{m_z}
\eeqa
To obtain a beta function from the above expression we follow the method proposed by Raghu {\it et al}\cite{raghu1} and further employed by Vafek and Wang\cite{wang1, wang2}. To this end we define the $g$ matrix $g_{\nu,\mu}(s,m_z) = \frac{U^2}{2^6} \sqrt{N_{\nu} N_{\mu}}v_{\nu, \mu}(s, m_z)$ from which we obtain

\beqa
g'_{\lambda,\lambda'}( s, m_z) &=&g_{\lambda, \lambda'}(m_z)\\ \nonumber & -&2  \ln(s) \sum_{\mu}  g_{\lambda, \mu}(m_z)g_{\mu,\lambda'}(m_z)
\eeqa
The matrix $v_{\nu, \mu}(s, m_z)$ is Hermitian (as we will show an a later section of this appendix) and thus so is $g$. We can then diagonalize $g$ as follows
\beq
g_{\nu, \mu}(m_z) = \sum_{i} \lambda^{m_z}_{i}(1) \psi^*_{i,\nu} \psi_{i,\mu}
\eeq
where $\lambda^i(1)$ are the eigenvalues of $g_{\nu, \mu}(m_z)$ and $ \psi_{i,\mu}$ is a vector whose columns are the (complete and orthonormal) eigenvectors of $g$. Using this in the above we then have
\beqa
g'_{\lambda,\lambda'}( s, m_z) &=& \sum_{i} \psi_{i, \lambda}^* \left(\lambda^{m_z}_i(1)-2  \ln(s) (\lambda^{m_z}_i(1))^2\right)\psi_{i, \lambda'}
\eeqa
The above tells that $g'_{\lambda,\lambda'}( s, m_z) $ is also diagonalized by this transformation and we obtain the result for the evolution of the eigenvalues of $g'_{\lambda,\lambda'}( m_z) $ under renormalization
\beqa
\lambda^{m_z}_i(s) &=& \lambda^{m_z}_i(1)-2  \ln(s) (\lambda^{m_z}_i(1))^2
\eeqa
The beta function for $\lambda_i$ is now obtained\cite{raghu1, wang1,wang2}  by taking the derivative of the above with respect to $\ln(s)$ which gives
\beqa
\frac{d \lambda^{m_z}_i(s)}{d \ln(s)}  &=&-2   (\lambda^{m_z}_i(1))^2 = -2   (\lambda^{m_z}_i(s))^2
\eeqa
where the second equality holds up to $\mathcal{O}(U^4)$. The solution to the above beta function is then
\beq
 \lambda^{m_z}_i(s) = \frac{ \lambda^{m_z}_i(1)}{1+2 \lambda^{m_z}_i(1)\ln(s)}
\eeq

Next we move on to the flow of the $m_z=0$ channel. The equation for the renormalized $v'_{\lambda,\lambda'}( s, m_z=0)$ is much more complicated than its $m_z\ne0$ counterpart. In order to obtain a flow equation one defines the matrix $g_{\lambda, \lambda'}(0) = \sqrt{N_{\lambda} N_{\lambda'}}\left(\frac{U F_{\lambda} F_{\lambda'}}{4}+\frac{U^2}{2^6} v_{\lambda, \lambda'}(s, 0)\right)$ along with $g'_{\lambda, \lambda'}(s,0) = \sqrt{N_{\lambda} N_{\lambda'}}v'_{\lambda,\lambda'}( s, 0) $ in order to find the following result which is valid up to $\mathcal{O}(U^4)$
\beqa
g'_{\lambda,\lambda'}(s, 0)  &=& g_{\lambda, \lambda'}(0) -2 \ln(s)\sum_{\mu} g_{\lambda, \mu}(0)  g_{\mu, \lambda'}(0)   \\ \nonumber &+&4 \ln^2(s)\sum_{\mu, \nu} g_{\lambda, \mu}(0)  g_{\mu, \nu}(0)   g_{\nu, \lambda'}(0)  \\ \nonumber & -&8 \ln^3(s)\sum_{\mu, \nu, \rho} g_{\lambda, \mu} (0) g_{\mu, \nu}  (0) g_{\nu, \rho} (0)  g_{\rho, \lambda'} (0)
\eeqa
We now diagonalize  $g_{\lambda,\lambda'}(0)$ and find that in the new basis $g'_{\lambda,\lambda'}(s, 0)$ is diagonal as well. This gives the result for the eigenvalues
\beqa
\lambda'_{i}(s)  &=& \lambda^{0}_{i}(1)\sum_{n}  (-2 \ln(s)( \lambda^{0}_{i}(1) ))^n   \\ \nonumber
&\simeq& \frac{\lambda_{i}(0)}{1+2\lambda_{i}(0)\ln(s)}
\eeqa
Taking the derivative of the above gives the beta function
\beqa
\frac{d \lambda^{0}_i(s)}{d \ln(s)}  &=& -2   (\lambda^{0}_i(s))^2
\eeqa
the same as that for the $m_z\ne0$ result but with a different initial condition\cite{wang1,wang2}.

\subsection{Hermiticity of $v_{\nu, \mu}(s, m_z)$}

It is tedious but straightforward to show that $\mathcal{F}^*_{\mu,\nu, \lambda_5,\lambda_6}(k_f^{\mu},k_f^{\nu}, \pv, \phi)=\mathcal{F}_{\nu,\mu, \lambda_5,\lambda_6}(k_f^{\nu},k_f^{\mu}, \pv,- \phi)$. From this property it follows that
 \beqa
(v^{m_z}_{\mu,\nu})^*&=&  \sum_{\lambda_5, \lambda_6}\int_0^{2\pi} \frac{d\phi}{2\pi}e^{im_z\phi}\int \frac{d^2\pv}{(2\pi)^2}\\ \nonumber &\times&\left(   \frac{n_F(E_{ \lambda_5}(p))-n_F(E_{ \lambda_6}(q))}{E_{ \lambda_5}(p)-E_{ \lambda_6}(q)} \right)\mathcal{F}_{\nu,\mu, \lambda_5,\lambda_6}(k_f^{\nu},k_f^{\mu}, \pv,- \phi)
\eeqa
We can then send $\phi\to-\phi$ and then, as everything in the integrand depends on $\phi$ through either $e^{i\phi}$ or $\cos(\phi)$ we can shift $\phi\to\phi+2\pi$. From this we immediately obtain $(v^{m_z}_{\mu,\nu})^*=v^{m_z}_{\nu,\mu}$. This is important because it ensures that $v^{m_z}_{\nu,\mu}$ and thus the $g's$ defined above can be diagonalized by a unitary transformation.
\end{widetext}
\end{document}